\documentclass[aps,prl,10pt,twocolumn]{revtex4-2}
\pdfoutput=1

\usepackage{amssymb}
\usepackage{amsfonts}
\usepackage{amsmath}
\usepackage{graphicx}%
\providecommand{\U}[1]{\protect\rule{.1in}{.1in}}
\begin{document}
\title{\textit{Ab initio} description of bcc iron with correlation matrix
renormalization theory }
\author{Jun Liu$^{a}$, Yongxin Yao$^{a,b}$, Vladimir Antropov$^{a,b}$, Kai-Ming
Ho$^{a,b}$, Cai-Zhuang Wang$^{a,b}$}
\affiliation{$^{a}$Ames Laboratory --USDOE, Iowa State University, Ames, IA 50011}
\affiliation{$^{b}$Department of Physics and Astronomy, Iowa State University, Ames, IA 50011}
\pacs{PACS number}

\begin{abstract}
We applied the \textit{ab initio} spin-polarized Correlation Matrix
Renormalization Theory (CMRT) to the ferromagnetic state of the bulk BCC iron.
We showed that it was capable of reproducing the equilibrium physical
properties and the pressure-volume curve in good comparison with experiments.
We then focused on the analysis of its local electronic correlations. By
exploiting different local fluctuation-related physical quantities as measures
of electronic correlation within target orbits, we elucidated the different
roles of $t_{2g}$ and $e_{g}$ states in both spin channels and presented
compelling evidence to showcase this distinction in their electronic correlation.

\end{abstract}
\date{6/26/2023}
\maketitle

%to convert between SWP and latex, the following revisions/care needed:
%(0) uncomment \documentclass[aps,prl,10pt,twocolumn]{revtex4-2}
%(0a) comment %\documentclass[preprint,prb]{revtex4}
%(1) comment/uncomment widetext
%(2) comment/uncomment natheight
%(3) comment/uncomment natwidth
%(3a) ignore l.157 {E-V.png} type error/pause
%(4) uncomment \bibliographystyle{apsrev4-2}
%(4a) comment %\bibliographystyle{plain}

\section{Introduction}

Iron, a prototypical magnetic material, is integral to our daily lives.
Experimental studies have ascertained that its low-temperature ground state,
the $\alpha$-Fe phase, exhibits a bcc crystal structure with an equilibrium
lattice volume of 11.7\AA
%TCIMACRO{\U{b3} }%
%BeginExpansion
${{}^3}$
%EndExpansion
(equivalent to lattice constant $a$=2.86\AA ) and a bulk modulus of
168GPa\cite{IntroToSolidStatePhys_Kittel2004}. As a ferromagnetic substance,
it possesses an ordered spin magnetic moment of 2.13$\mu_{B}$ and orbital
magnetic moment of 0.08$\mu_{B}$\cite{LandoltBornstein2017}. The system
displays discernible electronic correlation. Specifically, an effective local
Hubbard interaction U for 3d electrons in iron has been identified within a
range of 1$\sim$3eV, and a definitive ratio of $U/W\simeq0.2$ was established,
with $W$ representing the bandwidth of 3d
states\cite{Antonides1977_Auger_Ueff_Fe,Yin1977_Auger_delocal_Ueff_PhysRevB.15.2974,Chandesris1983_ARPES_Fe-Satellite}%
. This observation was later corroborated by a theoretical study coming up
with a close $U/W$
ratio\cite{Sasioglu2011_cLDA_ScreenU_fullyScreenU_PhysRevB.83.121101}. Such
characteristics were further evidenced in various experimental outcomes that
diverge from their mean field-like theoretical predictions and
interpretations\cite{Gutierrez1997_ARPES_Fe-Satellite_PhysRevB.56.1111,Chandesris1983_ARPES_Fe-Satellite}%
. Presently, both experimental and theoretical efforts have categorized
$\alpha$-Fe as a local moment system with a great tendency towards
itinerancy\cite{Wohlfarth1980_Fe-Ni_co_basic}. But a consensus is yet to be
reached on the underlying physical mechanism on the formation of the strong
ferromagnetism in $\alpha$%
-Fe\cite{Staunton1984_DLM_JMMM.45.15,Stearns1973_95Pecent_LocalMom_PRB,Singh1991_GGA_Fe_DOSatEf,Katanin2010_LDA-DMFT_U2.3J0.9_PhysRevB.81.045117}%
.

Density Functional Theory (DFT), including Local Spin Density Approximation
(LSDA) and its Generalized Gradient Approximation (GGA), has been applied to
$\alpha$-Fe to understand its peculiar physical properties from a microscopic
perspective. LSDA's predictions deviated from experimental findings and
suggested a notably reduced equilibrium lattice constant for the ferromagnetic
ground state of $\alpha$-Fe\cite{Wang1985_LSDA_wrongGS}. Adjusting this
discrepancy involves enhancing the kinetic energy via nonlocal charge density
variations and employing compatible exchange-correlation functionals akin to
GGA. These modifications yielded commendably accurate depictions concerning
the right ferromagnetic ground state and its innate
properties\cite{Singh1991_GGA_Fe_DOSatEf,Stixrude1994_GGA_Fe_PV}. Broadly, DFT
furnishes a reasonable portrayal of $\alpha$-Fe, including its energy ground
state and quasiparticle
characteristics\cite{Kresse1999_PRB_Fe_LDA-PAW,Stixrude1994_GGA_Fe_PV,schafer2005_ARPES_FSContour}%
. Specifically, it validates the Stoner mechanism for the emergence of
spontaneous ferromagnetism in BCC Fe\cite{Gunnarsson1976_StonerCrit-LDA_JPF}.
Other weakly interacting techniques, for example, GW
approximation\cite{Yamasaki2003_GW} and quasiparticle self-consistent
GW\cite{Kotani2007_QSGW,Sponza2017_QSGW_Fe-Ni_PhysRevB.95.041112}, have also
been applied to the system, purporting enhanced efficacy relative to GGA. A
semi-\textit{ab initio} Hartree-Fock (HF) calculation, where local and
nonlocal interaction operators were separately scaled, was also reported to
have produced A quite consistent bandstructure as
DFT\cite{Barreteau2004_semiHF_Good-BandStr_Fe-Co-Ni_PhysRevB.69.064432}.
Nevertheless, there is room for further refinement to illuminate the subtle
aspects of $\alpha$-Fe like local moment formation and competition between
localized and itinerant electrons, and to bridge the gap between theory and
experiments, notably through addressing both local and nonlocal electronic
correlations\cite{Belozerov2014_LDA-DMFT_U4J0.9_UJvalue,Sanchez2009_ARPES_LDA-DMFT_U1.5J0.9_PhysRevLett.103.267203,Lichtenstein2001_LDA-DMFT_U2.3J0.9,Pourovskii2014_LDA-DMFT_U4.3J1.0_EV-PV_PhysRevB.90.155120,Sponza2017_QSGW_Fe-Ni_PhysRevB.95.041112}%
.

Advanced \textit{ab initio} techniques, specifically designed to treat local
electronic correlation, have been employed to investigate the BCC iron system.
Notable methods included
LDA+U\cite{Cococcioni2005_LDAU_U2forFe_PhysRevB.71.035105}, LDA+Dynamic Mean
Field Theory
(LDA+DMFT)\cite{Katsnelson1999_DMFT-FLEX_U2.3J0.9,Lichtenstein2001_LDA-DMFT_U2.3J0.9,Katanin2010_LDA-DMFT_U2.3J0.9_PhysRevB.81.045117,Anisimov2012_LDA-DMFT_U2.3J0.9_PhysRevB.86.035152}
and LDA+Gutzwiller
(LDA+G)\cite{Deng2008_GLDA_FeU7J1,Borghi2014_PRB_GLDA_U2.5J1.2,Schickling2016_GLDA-U9J0.54}%
. While LDA+DMFT is considered the state-of-the-art\textit{ ab initio} method,
it is also computationally demanding. It was shown to improve the agreement
between theory and experiment, including very subtle aspects on quasiparticle
properties like broadening of quasiparticle
spectra\cite{Katsnelson1999_DMFT-FLEX_U2.3J0.9}, local spin
splitting\cite{Katanin2010_LDA-DMFT_U2.3J0.9_PhysRevB.81.045117,Anisimov2012_LDA-DMFT_U2.3J0.9_PhysRevB.86.035152}
and the emergence of satellite
subband\cite{Grechnev2007_LDA-DMFT_QSSpectr_U2.3J0.9_PhysRevB.76.035107}.
Specifically, it gave numerical evidence on the distinct nature of the
$t_{2g}$ and $e_{g}$ states in
electronic\cite{Katanin2010_LDA-DMFT_U2.3J0.9_PhysRevB.81.045117} as well as
magnetic contexts\cite{Kvashnin2015_J2_LDA-DMFT_U2.3J0.9_PhysRevB.91.125133},
and ascribed local moment mainly to $e_{g}$
electrons\cite{Katanin2010_LDA-DMFT_U2.3J0.9_PhysRevB.81.045117}. LDA+G can be
regarded as a simplified and accelerated version of LDA+DMFT with a different
definition of the Baym-Kadanoff functional within the conserving
approximation\cite{Lanata2015_LDA-G_RelatedTo_LDA-DMFT_PhysRevX.5.011008}. It
made specific physical observations based on its output and produced
information on quasiparticle dispersion. The engaged treatment OF local
electronic interactions helped introduce new interpretations towards
ferromagnetism from DFT methods\cite{Borghi2014_PRB_GLDA_U2.5J1.2}. However,
The notable challenge with these methods is the variability in defining
effective Hubbard $U$ and exchange $J$ parameters. These parameters are
essential for outlining screened local electronic interactions. They could
differ significantly across separate implementations and were often calibrated
to align with certain experimental
data\cite{Deng2008_GLDA_FeU7J1,Borghi2014_PRB_GLDA_U2.5J1.2,Schickling2016_GLDA-U9J0.54,Sanchez2009_ARPES_LDA-DMFT_U1.5J0.9_PhysRevLett.103.267203,Belozerov2014_LDA-DMFT_U4J0.9_UJvalue}%
. Specifically, the $U$ value can range from 2eV to 9eV, and $J$ between 0.5eV
and 1.2eV, a considerable spread for similar \textit{ab initio} techniques.
Nevertheless, there were reassuring studies indicating that magnetic
properties are more influenced by $J$ than $U$%
\cite{Schickling2016_GLDA-U9J0.54,Belozerov2014_LDA-DMFT_U4J0.9_UJvalue}.

DFT and its embedding methods, including LDA+U, LDA+G, and LDA+DMFT mentioned
above, enriched our knowledge for a better understanding of the microscopic
origin of the ferromagnetism in the bulk bcc iron system by analyzing physical
quantities coming out of the calculations and confirmed the importance of the
role local electronic correlation plays in producing a more accurate theory to
meet experiments. Local physical quantities analyzed include local
self-energy, spectral function, and spin-spin susceptibilities mainly produced
in
LDA+DMFT\cite{Lichtenstein2001_LDA-DMFT_U2.3J0.9,Katanin2010_LDA-DMFT_U2.3J0.9_PhysRevB.81.045117}%
, local orbit occupation and mass renormalization
factor\cite{Borghi2014_PRB_GLDA_U2.5J1.2}, and local charge(spin)
distribution\cite{Schickling2016_GLDA-U9J0.54}. They have provided direct
evidence on existence of local moment, asymmetry between $t_{2g}$ and $e_{g}$
states, and notable influence from electronic correlation. In this work, we
aim to delve deeper into some of these subjects, employing data from the
recently introduced \textit{ab initio} method, Correlation Matrix
Renormalization Theory
(CMRT)\cite{CMRsr2016_JCTC.12.4806,CMRsr2018_PhysRevB.97.075142,CMRsr2020_JPCM_33.095902_sp_formalism}%
. Uniquely, CMRT utilizes Hartree-Fock (HF) rather than DFT for the
foundational single-particle effective Hamiltonian. A strength of integrating
HF into CMRT is its direct engagement with term-wise bare Coulomb
interactions, eliminating the need for adjustable $U,J$ energy parameters and
double counting choices, and avoiding self-interaction complications. However,
this approach also has drawbacks: HF offers a less realistic quasiparticle
foundation for CMRT. Therefore, ensuring that the many-body screening effects
are properly incorporated within CMRT is essential. We thus assessed the total
energy of the system and compared the derived pressure-volume curve to
experimental data to ensure they are closely aligned, a necessary step for
CMRT to proceed further. We then devised a series of correlation metrics to
discern distinct roles of $t_{2g}$ and $e_{g}$ states across spin channels.

\section{Methods}

CMRT is a fully \textit{ab initio} variational theory specifically tailored
for strongly correlated electron systems utilizing a multiband Gutzwiller
wavefunction as its trial state\cite{CMRsr2020_JPCM_33.095902_sp_formalism}.
Notably, in the context of transition metal systems, CMRT offers a cohesive
framework that accommodates both itinerant and localized electrons within the
same electronic structure calculation, akin to DFT-embedded correlated
\textit{ab initio} methodologies\cite{Borghi2014_PRB_GLDA_U2.5J1.2}.

For a periodic bulk system with one atom per unit cell, the CMRT ground state
total energy is
%%if using revtex from APS, then uncomment the corresponding structure
\begin{widetext}%
\begin{equation}
E_{total}=\sum_{\substack{ij\\\alpha\beta,\sigma}}\tilde{t}_{i\alpha
,j\beta;\sigma}\left\langle c_{i\alpha\sigma}^{\dag}c_{j\beta\sigma
}\right\rangle +\dfrac{1}{2}\sum_{\substack{ijkl\\\alpha\beta\gamma
\delta,\sigma\sigma^{\prime}}}\tilde{U}_{ijkl;\sigma\sigma^{\prime}}%
^{\alpha\beta\gamma\delta}\left(  \left\langle c_{i\alpha\sigma}^{\dag
}c_{k\gamma\sigma}\right\rangle \left\langle c_{j\beta\sigma^{\prime}}^{\dag
}c_{l\delta\sigma^{\prime}}\right\rangle -\delta_{\sigma\sigma^{\prime}%
}\left\langle c_{i\alpha\sigma}^{\dag}c_{l\delta\sigma^{\prime}}\right\rangle
\left\langle c_{j\beta\sigma^{\prime}}^{\dag}c_{k\gamma\sigma}\right\rangle
\right)  +E_{local} \label{ecmrB}%
\end{equation}
%%uncomment the following line if using revtex from APS
\end{widetext}
with the local energy, $E_{local},$ expressed as
\begin{equation}
E_{local}=\sum_{i}\sum_{\Gamma}\tilde{E}_{i\Gamma}\left(  p_{i\Gamma
}-p_{i\Gamma_{0}}\right)  \label{ecmrL}%
\end{equation}
and the dressed hopping and two-body interactions are defined as%
\begin{align}
\tilde{t}_{i\alpha,j\beta;\sigma}  &  =t_{i\alpha,j\beta}+\dfrac{N_{e}}%
{2}\lambda_{ijji;\sigma\sigma}^{\alpha\beta\beta\alpha}\\
\tilde{U}_{ijkl;\sigma\sigma^{\prime}}^{\alpha\beta\gamma\delta}  &
=U_{ijkl}^{\alpha\beta\gamma\delta}-\lambda_{ijkl;\sigma\sigma^{\prime}%
}^{\alpha\beta\gamma\delta}%
\end{align}
Here $i,j,k,l$ represent site indices, $\alpha,\beta,\gamma,\delta$ are
orbital indices, and $\sigma,\sigma^{\prime}$ correspond to spin indices.
$\Gamma$ denotes Fock states in the occupation number representation of local
correlated orbitals on each atom in the unit cell, while $N_{e}$ is the
system's electron count per unit cell. The energy parameters, $t_{i\alpha
,j\beta}$ and $U_{ijkl}^{\alpha\beta\gamma\delta}$ are the bare hopping and
Coulomb integrals, respectively. The sum rule correction coefficient,
$\lambda_{ijkl;\sigma\sigma^{\prime}}^{\alpha\beta\gamma\delta},$ is
introduced in CMRT to specifically enhance the accuracy of the total energy
calculation. $\tilde{E}_{i\Gamma}$ is the Fock state eigenvalues of the
dressed local correlated Hamiltonian on each site.

The initial two terms in Eq. \ref{ecmrB} yield the expectation value of the
dressed lattice Hamiltonian under CMRT, where the expectation values of
two-body operators expand following Wick's theorem in terms of one-particle
density matrices, which is defined as%
\begin{align}
\left\langle c_{i\alpha\sigma}^{\dag}c_{i\beta\sigma}\right\rangle  &
=f\left(  z_{\alpha\sigma}\right)  f\left(  z_{\beta\sigma}\right)
\left\langle c_{i\alpha\sigma}^{\dag}c_{i\beta\sigma}\right\rangle
_{0}\nonumber\\
&  +\left[  1-\delta_{\alpha\beta}f^{2}\left(  z_{\alpha\sigma}\right)
\right]  \bar{n}_{i\alpha\sigma}%
\end{align}
Here, $z_{\alpha\sigma}$ represents the Gutzwiller renormalization factor
while $\left\langle \ldots\right\rangle _{0}$ indicates the one-particle
non-interacting density matrix and $\bar{n}_{i\alpha\sigma}$ the local
electronic occupation of state $\alpha$. The function $f\left(  z_{\alpha
\sigma}\right)  $ is integrated to ensure CMRT aligns with the solution of an
exactly solvable model\cite{CMRsr2018_PhysRevB.97.075142} under certain
conditions. The third term in Eq. \ref{ecmrB} is essential for preserving
dominant local physics in CMRT by rigorously expressing the local correlated
energy through the variational parameter $p_{i\Gamma}.$ This parameter
$p_{i\Gamma}$ denotes the occupational probability of Fock state $\Gamma$
spanned by the correlated atomic orbits at site $i.$ The non-interacting
counterpart, $p_{i\Gamma_{0}},$ denotes the same quantity evaluated with the
mean field approximation and correlates with the local energy components
already assessed in the initial two terms of Eq. \ref{ecmrB}. The underlying
local correlated Hamiltonian behind the third energy term of Eq. \ref{ecmrB}
encompasses primary two-body Hubbard-type Coulomb interaction terms dominating
local spin and charge interactions. Its exact treatment particularly helps
preserve intrinsic local spin and charge fluctuation effects and generate
local magnetic moments. The Hund's coupling exchange interaction terms, which
are believed to be physically relevant for bcc
iron\cite{UJ-Depnd_Belozerov_JPCM2014,Katanin2010_LDA-DMFT_U2.3J0.9_PhysRevB.81.045117}%
, are approached in a mean field way in CMRT.

The sum rule correction coefficients, provisionally represented as
\begin{equation}
\lambda_{ijkl;\sigma\sigma^{\prime}}^{\alpha\beta\gamma\delta}=\lambda
_{i\sigma}^{\alpha}\delta_{ik}\delta_{jl}\left(  1-\delta_{ij}\right)
\delta_{\alpha\gamma}\delta_{\beta\delta}, \label{sumrule}%
\end{equation}
are integrated explicitly into CMRT to aid in counteracting errors associated
with the Fock terms in Eq. \ref{ecmrB}. These terms constitute a significant
error source of CMRT. The sum rule correction coefficients serve to
redistribute non-local Coulomb interactions onto local sites, thus further
refining total energy by exactly treating these local interactions. The
central term, $\lambda_{i\sigma}^{\alpha},$ in Eq. \ref{sumrule} for each
correlated orbit is tested out in this work for magnetic systems. Its optimal
functional form is determined following the logic of cancellation of
inter-site Fock contributions and is identified as
\begin{equation}
\lambda_{i}^{\alpha}=\dfrac{\sum_{\sigma^{\prime}}\left[  \sum_{j\neq i}%
\sum_{\beta}U_{ijij}^{\alpha\beta\alpha\beta}\left\vert \left\langle
c_{i\alpha,\sigma^{\prime}}^{\dag}c_{j\beta,\sigma^{\prime}}\right\rangle
\right\vert ^{2}\right]  }{\sum_{\sigma^{\prime}}\left[  \sum_{j\neq i}%
\sum_{\beta}\left\vert \left\langle c_{i\alpha,\sigma^{\prime}}^{\dag
}c_{j\beta,\sigma^{\prime}}\right\rangle \right\vert ^{2}\right]  }
\label{sumrule-1}%
\end{equation}
One reassuring aspect of the above definition is the spin-independent nature
of the term, which aligns with the system's bare \textit{ab initio}
Hamiltonian. There, the energy coefficients of one-body and two-body operators
are all spin-independent. Thus, whatever magnetization produced in CMRT is a
genuine characteristic of the system but not endowed by certain pre-defined
energy parameters.

The variational minimization of the CMRT total energy, as given by Eq.
\ref{ecmrB}, yields a set of Gutzwiller
equations\cite{CMRsr2020_JPCM_33.095902_sp_formalism}. These are
self-consistently solved to reach the optimal solution for the target system.
For weakly correlated lattice systems, the volume-dependent total energy and
related physical quantities produced by CMRT have been found to align closely
with experimental results\cite{CMRsr2020_JPCM_33.095902_sp_formalism}. In the
realm of strongly correlated systems, CMRT has demonstrated its prowess in
capturing the correlated nature of 4f electrons in fcc Ce and fcc Pr
\cite{Jun2021_CMRT-Ce_PhysRevB.104.L081113}. By interfacing with the
Hartree-Fock (HF) module of Vienna Ab Initio Simulation Package (VASP)
\cite{VASP_PhysRevB.54.11169}, CMRT has been efficiently implemented with the
QUAsi-atomic Minimal Basis set Orbitals (QUAMBO) basis set
\cite{QUAMBO_PhysRevB.78.245112}. Its computational speed mirrors that of a
minimal basis HF calculation
\cite{CMRsr2020_JPCM_33.095902_sp_formalism,Jun2021_CMRT-Ce_PhysRevB.104.L081113}%
, marking a significant performance gain over the more time-consuming Quantum
Monte Carlo methods. Specifically for this work, a plane-wave basis set was
constructed in VASP with the default energy cutoff prescribed by the
pseudopotential of Fe. Brillouin zone sampling was facilitated with VASP using
an automatically generated K-point grid maintaining a $R_{k}$ length of 40
($R_{k}$ =40), which amounts to a $20\times20\times20$ uniform mesh at the
experimental lattice constant. The local QUAMBO basis set of 3d4s4p states are
projected from the LDA wavefunction preserving the low-energy LDA spectrum up
to 1eV above the LDA Fermi energy. These localized orbits define the tight
binding Hamiltonian and the bare Coulomb interactions.%

\begin{figure*}[ptb]%
\centering
\includegraphics[
%natheight=4.719300in,
%natwidth=5.899700in,
height=2.5867in,
width=3.2309in
]%
{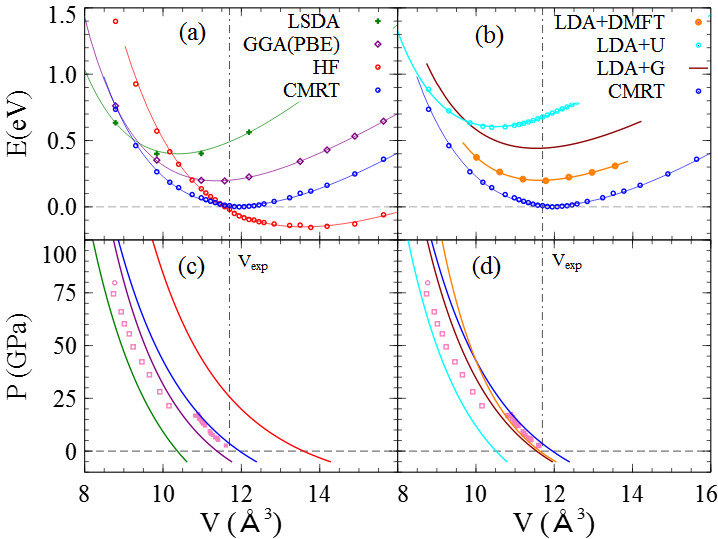}%
\caption{Energy versus Volume (E-V) and Pressure versus Volume (P-V) curves
for ferromagnetic bcc iron calculated with different \textit{ab initio}
methods. Plot (a) compares CMRT against weakly interacting methods including
LSDA, GGA(PBE) and HF while plot (b) compiles strongly correlated methods,
including LDA+DMFT ($U$=4.3eV, $J$%
=1.0eV)\cite{Pourovskii2014_LDA-DMFT_U4.3J1.0_EV-PV_PhysRevB.90.155120}, LDA+U
($U$=2.2eV, $J$=1.0eV), LDA+G ($U$=2.2eV, $J$%
=1.0eV)\cite{Deng2008_GLDA_FeU7J1} and CMRT. The corresponding P-V curves are
depicted in solid lines in plots (c) and (d). LSDA, PBE, and LDA+U are
evaluated with VASP at an automatic K grid of $R_{k}$=40. HF used the
identical QUAMBO basis set and K-grid as CMRT. Experimental measurements are
symbolized as follows, : solid squares for bcc $\alpha$-Fe phase and empty
squares for hcp $\varepsilon$-Fe
phase\cite{Stixrude1994_GGA_Fe_PV,Dewaele2015_a-e_marten}. Vertical
adjustments have been made for all the energy curves for a clearer view.
Specifically, HF energy is downshifted an extra 5.5eV with respect to CMRT
energy in the figure. Vertical dash dotted lines mark the experimental
equilibrium volume of the ferromagnetic BCC Fe lattice.}%
\label{E-V}%
\end{figure*}
%EndExpansion

\section{Results}

\subsection{Total energy and its related physical quantities}

\begin{table*}[th]
\caption{BM-EOS fitted equilibrium lattice constant $a_{0}$, bulk modulus
$B_{0}$ and the calculated spin magnetic moment $M$ of ferromagnetic BCC Fe
obtained by various calculation methods are in comparison with experimental
measurements. HF, LDA+U, and CMRT data were calculated in this work, while
other data are sourced or adapted from relevant literature:
Exp\cite{Pourovskii2014_LDA-DMFT_U4.3J1.0_EV-PV_PhysRevB.90.155120}, LSDA and
GGA(PBE) \cite{Kresse1999_PRB_Fe_LDA-PAW},
LDA+DMFT\cite{Pourovskii2014_LDA-DMFT_U4.3J1.0_EV-PV_PhysRevB.90.155120},
LDA+G\cite{Deng2008_GLDA_FeU7J1}. Note that LDA+DMFT utilized $U$=4.3eV and
$J$=1.0eV to match the experimental spin magnetic moment value of 2.2$\mu_{B}$
at T=290K. Meanwhile, while there are varied parameter choices for LDA+G in
the literature, here it uses $U$=7eV and $J$=1.0eV. Also, note the spin
magnetic moment tabulated here is linked to the equilibrium lattice constant
specific to each method, and not pegged to the experimental lattice constant.}%
\label{tab1}
\centering
\begin{tabular}
[t]{lcccccccc}\hline
& Exp & HF & LSDA & PBE & LDA+U & LDA+G & LDA+DMFT & CMRT\\\hline
$a_{0}$ (\AA ) & 2.867 & 3.0 & 2.746 & 2.833 & 2.76 & 2.85 & 2.853 & 2.887\\
$B_{0}$ (GPa) & 172 & 115 & 245 & 169 & 207 & 160 & 168 & 165\\
$M$ ($\mu_{B}$) & 2.2 & 2.92 & 2.00 & 2.2 & 2.13 & 2.30 & 2.2 & 2.6\\\hline
&  &  &  &  &  &  &  & \\
&  &  &  &  &  &  &  &
\end{tabular}
\end{table*}

In the study of the ferromagnetic ground state of the bulk bcc Fe lattice,
energy versus volume (E-V) curves are collected and compared in panel (a) and
(b) of Fig \ref{E-V}. These curves contain results from several calculation
methods, including HF, LSDA, GGA(PBE), LDA+U, LDA+G, LDA+DMFT and CMRT. Both
HF and CMRT calculations share the same QUAMBO basis set, while LSDA, GGA(PBE)
and LDA+U are evaluated with plane-wave basis set in this work. GGA data are
cross-checked against the published results in Ref.
\cite{Stixrude1994_GGA_Fe_PV}. To complement the E-V curves, the pressure
versus volume (P-V) curves extracted from their Birch-Murnaghan Equation of
State (BM-EOS) \cite{Murnaghan_PNAS.30.244} fits are also showcased in panel
(c) and (d) of Fig \ref{E-V}, side by side with the experimental measurements,
while the accompanying fitted equilibrium volumes and bulk moduli as well as
the calculated magnetic moments are collected in Table \ref{tab1}. By
examining the intersection points of these curves with the volume axis, we can
discern the distribution of equilibrium volumes for each method in relation to
the experimental volume. This provides a clear illustration of the exemplary
performance of both the GGA and CMRT methods, which operate without the need
for adjustable energy parameters, and commendable outcomes of LDA+G and
LDA+DMFT with appropriate $U$, $J$ energy parameters adapted. The alignment
between the CMRT-generated data and experimental pressure-volume measurements
stands out. Specifically, CMRT demonstrates a closer resemblance to
experimental outcomes for the bcc iron phase when compared to GGA.
\begin{figure*}[ptb]%
\centering
\includegraphics[
%natheight=6.530200in,
%natwidth=4.249700in,
height=4.2454in,
width=2.7735in
]%
{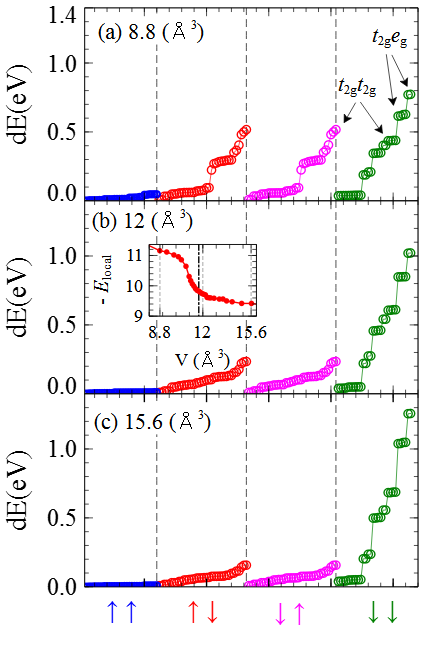}%
\caption{Displayed are the two-body energy components contributing to local
energy correction as defined in Eq. \ref{ecmrL} at three typical volumes of
the BCC iron lattice. Empty circles represent each two-body energy component
given by $dE=U_{iiii}^{\alpha\alpha\beta\beta}\left(  \bar{n}_{i\alpha\sigma
}\bar{n}_{i\beta\sigma^{\prime}}-\left\langle \hat{n}_{i\alpha\sigma}\hat
{n}_{i\beta\sigma^{\prime}}\right\rangle \right)  $ with $U_{iiii}%
^{\alpha\alpha\beta\beta}$ local bare Coulomb integral matching $\hat
{n}_{i\alpha\sigma}\hat{n}_{i\beta\sigma^{\prime}}$ operator, and are sorted
in ascending order on the x-axis for each spin-spin channel at a designated
volume. Consequently, symbols located at the same x-coordinate in different
volumes may not represent the same operator unless explicitly indicated.
Vertical dashed lines distinguish these channels with corresponding
consistently colored circles. The up arrow signifies majority spin, while the
down arrow indicates minority spin. Prominent local energy correction
components pointed with arrows are tagged as $t_{2g}e_{g}$ and $t_{2g}t_{2g},$
representing local two-body operators of $\hat{n}_{t_{2g},\sigma}\hat
{n}_{e_{g},\sigma}$ and $\hat{n}_{t_{2g},\sigma}\hat{n}_{t_{2g}^{\prime
},\sigma^{\prime}}$ respectively. The inset illustrates the volume dependence
of local energy correction induced by electronic correlation and expressed as
$-E_{local}$ in the CMRT energy expression of Eq. \ref{ecmrB}. The thick
dashed vertical line there marks the experimental equilibrium volume, while
the trio of dashed grid lines point out where energy components were sampled.
}%
\label{eng_2D}%
\end{figure*}
%EndExpansion

One might wonder how local energy corrections resulting from electronic
correlations might influence the total energy in CMRT calculations. This
particular contribution is encapsulated in $E_{local}$ as seen in Eq.
\ref{ecmrB}. As described by Eq. \ref{ecmrL}, $E_{local}$ encompasses
predominant energy terms arising from $\hat{n}_{i\alpha,\sigma}\hat{n}%
_{i\beta,\sigma^{\prime}}$ type of two-body operators, where $\alpha$ and
$\beta$ represent the set of local correlated orbits. This term delineates the
discrepancy between the strict expectation values and their corresponding mean
field values. Typically, each term in $E_{local}$ is negative, reflecting
diminished Coulomb interaction stemming from the presence of local electronic
repulsion. We've assigned an additional negative sign to these terms for a
clearer visualization in Fig. \ref{eng_2D}. General understanding might
suggest that local correlation energy gain amplifies with increasing volume
expansion. Yet, contrary to this notion, the inset of the figure displays a
different trend. The root of this behavior can be traced back to the terms
that most significantly influence $E_{local},$ as exemplified at three
distinct volumes across the experimental equilibrium volume. These individual
energy terms are segregated into separate spin-spin channels on the x-axis of
Fig. \ref{eng_2D}: $\uparrow\uparrow$ for majority-majority spin,
$\uparrow\downarrow$ for majority-minority spin, and so forth. A closer look
reveals that energy corrections from the majority-majority spin channel remain
minuscule across the considered terms following the x- axis. The
majority-minority spin channel flourishes while it contributes to $E_{local}$
at a reduced volume but diminishes rapidly beyond the experimental volume.
Conversely, the minority-minority spin channel possesses a handful of two-body
operators that notably amplify their contributions to $E_{local},$ indicating
a swift rise in electronic repulsion between specific states. The composite
energy correction trajectory, presented in the inset, unveils that the gains
from enhanced terms in the minority-minority spin channel fail to offset the
dwindling contributions from the majority-minority two-body terms.

\subsection{Local Orbital Occupations and Their Fluctuations}

\begin{figure*}[ptb]%
\centering
\includegraphics[
%natheight=4.350000in,
%natwidth=5.840100in,
height=2.4491in,
width=3.2811in
]%
{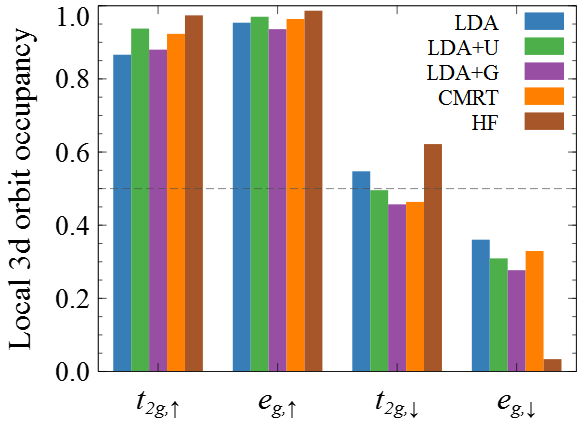}%
\caption{Local charge occupation of th $t_{2g}$ and $e_{g}$ states within the
3d orbit collected from HF, CMRT, LDA+U ($U$=2.3eV, $J$=0.9eV) and LDA+G
($U$=2.5eV, $J$=1.2eV) \cite{Borghi2014_PRB_GLDA_U2.5J1.2}$.$}%
\label{NCPHY}%
\end{figure*}

A comprehensive examination of the local physics is presented in the
ferromagnetic bcc iron lattice using the CMRT method. Fig \ref{NCPHY} gives
local orbital occupancies on the $t_{2g}$ and $e_{g}$ states of the 3d orbit
at a lattice volume of $11.94$\AA $^{3}$ (or $a=2.88$\AA ) across various
\textit{ab initio} methods. The orbital occupancies of CMRT align closely with
most methods except for HF. For example, using the same QUAMBO local orbit
basis set, both LDA and CMRT yield roughly 1.3 electrons in each of the
$t_{2g}$ and $e_{g}$ states though CMRT exhibits a slightly greater ordered
spin magnetic moment. On the other hand, a discrepancy in the HF orbital
occupancy is evident in the minority spin channel, where the $t_{2g}$ state
occupation significantly surpasses that of the $e_{g}$ state. This disparity
may indicate that the local 3d energy components dominate the HF total energy.
More details are provided in the discussion. The CMRT formalism, built upon
the HF method, incorporates electronic correlation effects through both
renormalizing effective single particle hoppings and rigorously treating local
two-body interactions. Such a procedure successfully reduces electron
occupancy in the majority spin channel and markedly redistributes electrons
between the $t_{2g}$ and $e_{g}$ states in the minority spin channel, yielding
more balanced orbital occupancies and tempering the pronouncedly high local
spin moment returned by HF.%
%TCIMACRO{\FRAME{ftbpFU}{3.2811in}{2.4491in}{0pt}{\Qcb{Local charge occupation
%of th $t_{2g}$ and $e_{g}$ states within the 3d orbit collected from HF, CMRT,
%LDA+U ($U$=2.3eV, $J$=0.9eV) and LDA+G ($U$=2.5eV, $J$=1.2eV)
%\cite{Borghi2014_PRB_GLDA_U2.5J1.2}$.$}}{\Qlb{NCPHY}}{ncphy.png}%
%{\special{ language "Scientific Word";  type "GRAPHIC";
%maintain-aspect-ratio TRUE;  display "USEDEF";  valid_file "F";
%width 3.2811in;  height 2.4491in;  depth 0pt;  original-width 5.8401in;
%original-height 4.35in;  cropleft "0";  croptop "1";  cropright "1";
%cropbottom "0";  filename 'NCPHY.png';file-properties "XNPEU";}} }%
%BeginExpansion
%EndExpansion

To delve deeper into fluctuations, we introduce a local pseudo-charge
correlator as
\begin{equation}
\chi_{i\alpha\sigma,i\beta\sigma^{\prime}}=\left\langle \hat{n}_{i\alpha
\sigma}\hat{n}_{i\beta\sigma^{\prime}}\right\rangle -\bar{n}_{i\alpha\sigma
}\bar{n}_{i\beta\sigma^{\prime}}\text{ for }\left(  \alpha\sigma\right)
\neq\left(  \beta\sigma^{\prime}\right)  \label{correlator}%
\end{equation}
This correlator serves as an insightful metric to gauge the electronic
correlation between two electronic states effectively capturing how one
electron's presence might influence another's motion. In essence, this
correlator quantifies the deviation in the likelihood of observing a specific
electron pair, $\left\langle \hat{n}_{i\alpha\sigma}\hat{n}_{i\beta
\sigma^{\prime}}\right\rangle $, which can be thoroughly evaluated within
CMRT, from a baseline uncorrelated value, $\bar{n}_{i\alpha\sigma}\bar
{n}_{i\beta\sigma^{\prime}}$. When the expectation value is evaluated with a
single Slater determinant ground state wavefunction, the result would yield
the Hartree term as the baseline value, and a much smaller Fock term if the
working basis set possesses the correct lattice and orbital symmetry. Thus,
this correlator would nearly vanish in a non-interacting system, as expected
for two electrons being uncorrelated.

Introduce local charge and spin (z component only) operators as $\hat{n}$ and
$\hat{S}_{z}$ and we can write down the local static charge and spin (z
component only) fluctuations, $\chi_{\hat{n}}$ and $\chi_{\hat{S}_{z}},$ as%
\begin{gather}
\hat{n}=\sum_{\alpha,\sigma}\hat{n}_{\alpha,\sigma}\Rightarrow\chi_{\hat{n}%
}=\left\langle \left(  \hat{n}-\bar{n}\right)  ^{2}\right\rangle \\
\hat{S}_{z}=\dfrac{1}{2}\sum_{\alpha,\sigma}\sigma\hat{n}_{\alpha,\sigma
}\Rightarrow\chi_{\hat{S}_{z}}=\left\langle \left(  \hat{S}_{z}-\bar{S}%
_{z}\right)  ^{2}\right\rangle
\end{gather}
with $\alpha$ indexing a set of local orbits and $\sigma=\pm1$ denoting
majority and minority spins, respectively. A simple algebra establishes the
following relationship between fluctuations and pseudo-charge correlator
\begin{equation}
4\chi_{\hat{S}_{z}}-\chi_{\hat{n}}=4\sum_{\alpha\beta}\left(  -\chi
_{\alpha\uparrow,\beta\downarrow}\right)  \label{fluc_corr}%
\end{equation}
Given a single orbit, the above equation provides a way to gain insights into
local double occupancy by taking the difference between the two fluctuations.

\begin{figure*}[ptb]%
\centering
\includegraphics[
%natheight=5.570300in,
%natwidth=8.230400in,
height=4.3301in,
width=6.3858in
]%
{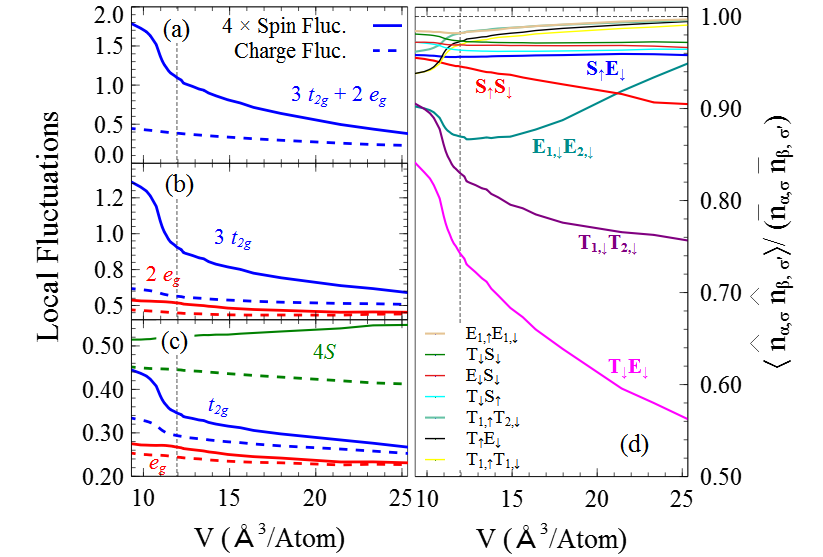}%
\caption{Left panels show local spin and charge fluctuations with different
sets of local states. Solid lines show four times spin fluctuation and dashed
lines charge fluctuation. Panel (a) selects all $t_{2g}$ and $e_{g}$ states to
define the spin and charge operators; panel (b) selects all $t_{2g}$ states as
well as all $e_{g}$ states while panel (c) picks up individual $4S,$ $t_{2g}$
and $e_{g}$ states to evaluate their spin and charge fluctuations. The right
panel (d) gives comparisons of $\left\langle \hat{n}_{\alpha\sigma}\hat
{n}_{\beta\sigma^{\prime}}\right\rangle $ normalized by $\bar{n}_{\alpha
\sigma}\bar{n}_{\beta\sigma^{\prime}}$ with $\bar{n}_{\alpha\sigma}=$
$\left\langle \hat{n}_{\alpha\sigma}\right\rangle $. The first five biggest
terms have their labels put nearby the curves with matching colors. In
contrast, the rest of the terms have their labels collected at the left-bottom
corner following roughly the magnitude ordering at a small lattice volume. The
curves are consistently labeled with two capital letters denoting a pair of
local orbits involved. Specifically, $T,E,S$ denote $t_{2g},$ $e_{g}$ and $4S$
states averaged over their degenerate states, respectively. If the same state
is involved in both local orbits, then each letter carries a number to
distinguish whether they are the same state. For instance, T$_{1,\uparrow}%
$T$_{2,\downarrow}$ denotes $\hat{n}_{\alpha\uparrow}\hat{n}_{\beta\downarrow
}$ with $\alpha,\beta\in\left\{  t_{2g}\right\}  $ but $\alpha\neq\beta,$
while T$_{1,\uparrow}$T$_{1,\downarrow}$ denotes $\hat{n}_{\alpha\uparrow}%
\hat{n}_{\alpha\downarrow}$ for $\alpha\in\left\{  t_{2g}\right\}  $, which is
basically the averaged double occupancy of an individual $t_{2g}$ state. The
spin index, $\sigma\in\left\{  \uparrow,\downarrow\right\}  ,$ denotes
majority or minority spins respectively. In both panels, the vertical dotted
lines show the CMRT equilibrium volume.}%
\label{MCFLUC}%
\end{figure*}

Fig. \ref{MCFLUC} compiles the spin and charge fluctuations from various sets
of local orbits and highlights the dominant pseudo-charge correlators. Panels
(a) and (b) dissect the fluctuations within all 3d orbits, and within $t_{2g}$
and $e_{g}$ states respectively. The principal variability in spin fluctuation
predominantly concerns the $t_{2g}$ states, especially at smaller lattice
volumes. Panel (c) provides a clearer perspective on the observation by
representing fluctuations for individual states. By noting that local
fluctuations of 4S state are not suppressible with increasing electronic
correlation, we might reliably classify 4S state to be weakly correlated.
Meanwhile, as volume increases, Panel (c) suggests that the $t_{2g}$ and
$e_{g}$ states exhibit weak correlation, as indicated by $\left\langle \hat
{n}_{i\alpha\uparrow}\hat{n}_{i\alpha\downarrow}\right\rangle \simeq\bar
{n}_{i\alpha\uparrow}\bar{n}_{i\alpha\downarrow}$ readily read out from the
diminishing difference between the spin and charge fluctuations and with help
of Eq. \ref{fluc_corr}. This weak correlation arises from the nearly filled 3d
orbits in the majority spin channel. The minority spin channel in the 3d
orbits, however, pose to be the chief contributor to local electronic
correlations. This observation stems from Fig. \ref{eng_2D} and is
corroborated by Panel (d) in Fig. \ref{MCFLUC}. This panel showcases
$\chi_{i\alpha\sigma,i\beta\sigma^{\prime}}$ adjusted by $\bar{n}%
_{i\alpha\sigma}\bar{n}_{i\beta\sigma^{\prime}}$ to account for variations in
orbital occupation. Such an approach can compare electronic correlations
across different state pairs, as is supported by two notable advantages.
First, all state pairs maintain their numerical alignment at one with the
non-interacting limit. Second, the visualization aptly highlights the few most
significant electronic correlations and pinpoints the state pairs that
generate them. These predominant correlations between $t_{2g}$ and $e_{g}$
could be the reason for their rebalanced occupations in CMRT which are
otherwise significantly skewed in the HF calculation shown in Fig. \ref{NCPHY}.%

%EndExpansion
%

%TCIMACRO{\FRAME{ftbpFU}{3.0087in}{2.6264in}{0pt}{\Qcb{Volume dependence of
%Gutzwiller renormalization factor, $f\left(  z_{\alpha\sigma}\right)  ,$ for
%both spin channels for selected local states. Blue curves are for $t_{2g},$
%red for $e_{g}$ and green for $4S$ states. Solid lines with an upward triangle
%represent the majority spin, while dashed lines with a downward triangle
%represent the minority spin. The dashed vertical line indicates the CMRT
%equilibrium volume with the ferromagnetic bcc iron lattice.}}{\Qlb{Z}%
%}{rout_r.png}{\special{ language "Scientific Word";  type "GRAPHIC";
%maintain-aspect-ratio TRUE;  display "USEDEF";  valid_file "F";
%width 3.0087in;  height 2.6264in;  depth 0pt;  original-width 6.2699in;
%original-height 5.4699in;  cropleft "0";  croptop "1";  cropright "1";
%cropbottom "0";  filename 'ROUT_r.png';file-properties "XNPEU";}} }%
%BeginExpansion
\begin{figure*}[ptb]%
\centering
\includegraphics[
%natheight=5.469900in,
%natwidth=6.269900in,
height=2.6264in,
width=3.0087in
]%
{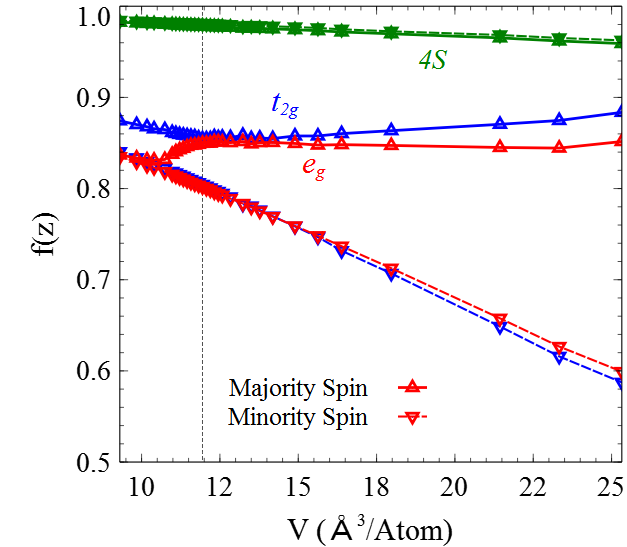}%
\caption{Volume dependence of Gutzwiller renormalization factor, $f\left(
z_{\alpha\sigma}\right)  ,$ for both spin channels for selected local states.
Blue curves are for $t_{2g},$ red for $e_{g}$ and green for $4S$ states. Solid
lines with an upward triangle represent the majority spin, while dashed lines
with a downward triangle represent the minority spin. The dashed vertical line
indicates the CMRT equilibrium volume with the ferromagnetic bcc iron
lattice.}%
\label{Z}%
\end{figure*}
%EndExpansion

\subsection{Normalized Local Charge Fluctuation analysis}%

%TCIMACRO{\FRAME{ftbpFU}{5.015in}{3.6625in}{0pt}{\Qcb{sNLCF results from both
%HF and CMRT methods are brought side by side in the plots. Labels on the top
%of the figure show different spin channels of sNLCF in the columns while
%labels to the right of the figure show the chosen local orbits for sNLCF
%calculation in each row. Specifically, $3d$ means that all 3d local orbitals
%of the prechosen spins are selected to define NLCF, $t_{2g}$ and $e_{g}$
%denote that a single specific state is chosen for sNLCF calculation. The inset
%in plot (b) shows lattice constant dependence of difference in sNLCF between
%HF and CMRT data with the dashed red a smooth fit of the difference to
%indicate the maximum. The dashed vertical line denotes the equilibrium lattice
%constant determined from CMRT. }}{\Qlb{sNLCF}}{nlcfd_r_norm_simple.png}%
%{\special{ language "Scientific Word";  type "GRAPHIC";
%maintain-aspect-ratio TRUE;  display "USEDEF";  valid_file "F";
%width 5.015in;  height 3.6625in;  depth 0pt;  original-width 7.6398in;
%original-height 5.5703in;  cropleft "0";  croptop "1";  cropright "1";
%cropbottom "0";  filename 'NLCFD_r_norm_simple.png';file-properties "XNPEU";}}
%}%
%BeginExpansion
\begin{figure*}[ptb]%
\centering
\includegraphics[
%natheight=5.570300in,
%natwidth=7.639800in,
height=3.6625in,
width=5.015in
]%
{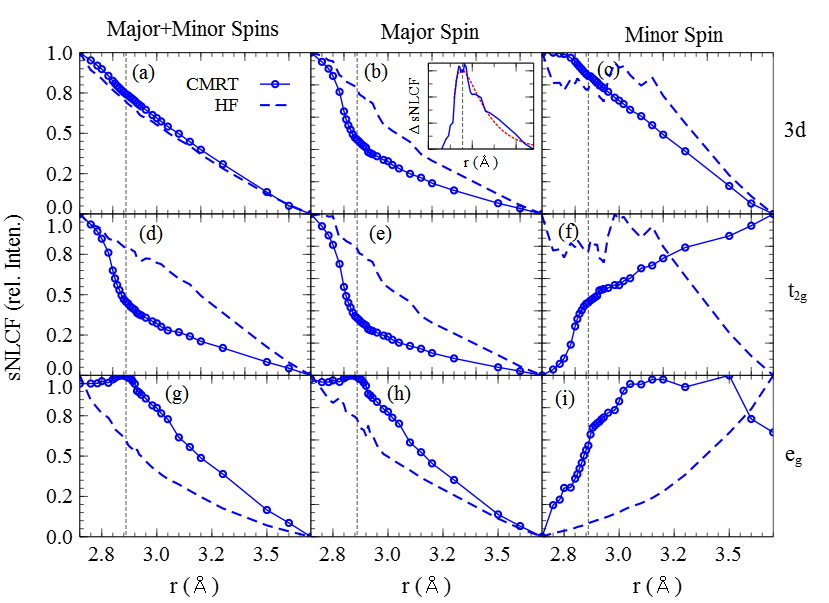}%
\caption{sNLCF results from both HF and CMRT methods are brought side by side
in the plots. Labels on the top of the figure show different spin channels of
sNLCF in the columns while labels to the right of the figure show the chosen
local orbits for sNLCF calculation in each row. Specifically, $3d$ means that
all 3d local orbitals of the prechosen spins are selected to define NLCF,
$t_{2g}$ and $e_{g}$ denote that a single specific state is chosen for sNLCF
calculation. The inset in plot (b) shows lattice constant dependence of
difference in sNLCF between HF and CMRT data with the dashed red a smooth fit
of the difference to indicate the maximum. The dashed vertical line denotes
the equilibrium lattice constant determined from CMRT. }%
\label{sNLCF}%
\end{figure*}
%EndExpansion

While the Gutzwiller renormalization prefactors for the correlated orbits
shown in Fig. \ref{Z} reveal some similarity between $t_{2g}$ and $e_{g}$
states in both spin channels, the difference might be explored through the
Normalized Local Charge Fluctuation (NLCF), defined as $\chi_{\hat{n}}/\bar
{n}^{2}$\cite{Jun2021_CMRT-Ce_PhysRevB.104.L081113}. We evaluate this metric
using CMRT and HF calculations, with HF serving as the reference for
electronic correlation. Notable deviations between CMRT and HF indicate
additional correlations captured by CMRT. For a balanced comparison, we
introduce a standardized NLCF (sNLCF). Given the non-comparability of
expectation values in the NLCF definition across methods, we adjusted their
range to fall between 0 and 1, considering unique constant shifts.

Fig \ref{sNLCF} contrasts sNLCF values from CMRT and HF across subsets of
local correlated orbits in a ferromagnetic bcc iron system. This figure
presents relative charge fluctuations across different choices of orbits
(rows) and spin channels (columns). The top row illustrates CMRT vs. HF for
all five 3d orbits, and the middle and bottom rows focus on comparisons for
individual $t_{2g}$ and $e_{g}$ states respectively. In interpreting Fig
\ref{sNLCF}, it's evident that different treatments in electronic correlation
between methods yield different sNLCF behaviors. Specifically, the majority
spin channel in the second column reveals HF's near-linear descent as
contrasted with CMRT's well-established curvatures. CMRT either further
suppresses or enhances charge fluctuations on top of HF in the $t_{2g}$ or
$e_{g}$ states for a better treatment of their electronic correlations. This
qualitative difference in $t_{2g}$ and $e_{g}$ treatment supports the distinct
correlation nature of both 3d states made in existing literature. The curve in
the inset of plot (b), resulting from the difference between HF and CMRT
there, peaks near the CMRT equilibrium volume. This might suggest a
predominant role of majority spin electrons in shaping the interatomic bonds
and the bulk bcc lattice structure.

\section{Discussion}

We demonstrated that CMRT can correctly predict both the energy versus volume
(E-V) and pressure versus volume (P-V) curves for the bulk BCC iron
ferromagnetic phase. Furthermore, it yields an equilibrium volume and bulk
modulus consistent with experimental findings, as illustrated in Fig.
\ref{E-V}. CMRT also produced other credible physical quantities like local
orbital renormalization prefactors and orbital occupations. All these suggest
that CMRT can capture the essential correlation physics inherent in the 3d
orbits of this system. These extra correlations built into CMRT aid in
redistributing the system's kinetic and potential energies, and orbital
fillings. While there was analysis indicating that changes in these energy
components correlate with the formation of ordered
moments\cite{Borghi2014_PRB_GLDA_U2.5J1.2}, we choose not to delve into such
intricacies here, given that this information might be method specific.
Meanwhile, CMRT predicts a local spin magnetic moment larger than experimental
measurements. The local state occupations depicted in Fig \ref{NCPHY} reveal
that HF-based CMRT still allocates more electrons to the majority spin channel
than LSDA/GGA, resulting in an exaggerated local spin magnetic moment.
Interestingly, local interaction enhanced LSDA methods, such as DFT+U and
DFT+G, display similar local state occupations as CMRT, even though they stem
from distinct theoretical backgrounds, namely LSDA and HF.

The local 3d occupation in HF significantly skews towards $t_{2g}$ states in
the minority spin channel compared to the other methods, as depicted in Fig.
\ref{NCPHY}. Generally speaking, a preferred occupation on $t_{2g}$ over
$e_{g}$ is consistent with the cubic crystal field splitting of 3d
orbits\cite{MagnetismCondMatt_blundell2001}. But, this skew in HF calculation
seems excessively pronounced. Insight into this phenomenon may be gleaned by
examining a simplified model of an isolated atom. This model replicates the
local electron filling pattern observed in the ferromagnetic iron state,
presupposing nearly fully filled 3d orbits in the majority spin channel and a
predetermined number of 3d electrons in the minority spin channel. We closely
observe $\hat{n}_{\alpha\sigma}\hat{n}_{\beta\sigma^{\prime}}$ type two-body
operators, with $\alpha,\beta\in\left\{  t_{2g},e_{g}\right\}  $, which are
dominant in the energy Hamiltonian and possess very close Coulomb energy
coefficients. The classical Coulomb potential energy pertinent to these
operators is expressed as follows
\begin{equation}
E_{p}\propto C_{3}^{2}\bar{n}_{t_{2g},\downarrow}\bar{n}_{t_{2g},\downarrow
}+6\bar{n}_{t_{2g},\downarrow}\bar{n}_{e_{g},\downarrow}+C_{2}^{2}\bar
{n}_{e_{g},\downarrow}\bar{n}_{e_{g},\downarrow} \label{localpot}%
\end{equation}
which might as well be thought of a mean field decomposition on $\hat
{n}_{\alpha\sigma}\hat{n}_{\beta\sigma^{\prime}}$ but having the Fock terms
dropped as quantum effects. In this equation, $\bar{n}_{t_{2g}\left(
e_{g}\right)  ,\downarrow}$ represents local orbital occupation in a $t_{2g}$
or $e_{g}$ state in the minority spin channel, while $C_{n}^{m}$ is the
standard binomial coefficient. Simple algebraic manipulation reveals three
notable cases\cite{HF-PotentialEnergy}. Two extremes, $\left(  \bar{n}%
_{t_{2g},\downarrow},0\right)  $ and $\left(  0,\bar{n}_{e_{g},\downarrow
}\right)  ,$ represent variable-bounded local minima separated by a potential
energy maximum, which defines the physically relevant third case holding equal
occupation in all 3d orbits for an isolated atom with nearly degenerate
orbits. Given this scenario, it is reasonable to hypothesize that the HF
solution likely corresponds to one of the two extreme cases in an effort to
minimize local potential energy. Confirmation of this hypothesis is obtained
by applying HF to the local energy Hamiltonian constructed at a reference site
on the BCC iron lattice with a unit cell volume of $25$\AA $^{3}$(or
$a_{0}=3.7$\AA ). The HF approach, contingent on specific initial orbital
occupations, readily converges to the two extreme cases with vanishing
occupation in either type of the 3d states. Comparing these solutions to the
actual HF solution for the BCC iron lattice reveals that the extra nonlocal
hoppings and interactions left out of the local energy Hamiltonian contribute
to electron transitions into the empty states. With local correlation effects
incorporated in the HF framework to establish CMRT -- which effectively
reduces local Coulomb interaction as showcased in Fig. \ref{MCFLUC} -- a
greater number of electrons continue to transit into the empty 3d states. This
results in a more balanced electron occupation among the 3d states, which
would otherwise be energetically discouraged by a local energy Hamiltonian as
seen through HF.

Based on the classical potential energy depicted in Eq. \ref{localpot} and the
different orbital occupations between HF and CMRT, two key observations are
made. Firstly, local correlation is crucial in reestablishing the correct
physical picture in the BCC iron lattice with CMRT. While correlation may
reduce nonlocal energy components through Gutzwiller renormalization, the
overarching effect is an enhanced nonlocal effect, ensuring a steady electron
flow into empty 3d states. Secondly, integrating the exchange-correlation
functional into DFT markedly enhances its efficacy, as evidenced here by a
correct depiction of the BCC iron lattice. Nevertheless, the similarity in
electronic behaviors yielded by both the classical Coulomb repulsion and HF
positions HF as a benchmark methodology in comparing treatment on electronic
correlation effects, which is purely quantum in nature. These insights might
be instrumental in resolving an inconsistent statement made in a QSGW
calculation\cite{Sponza2017_QSGW_Fe-Ni_PhysRevB.95.041112} stating that local
physics is not relevant for describing BCC iron lattice by taking DFT as its reference.

The local correlated energy, $E_{local},$ as defined in Eq. \ref{ecmrL}
encapsulates the effect of correlation on the electronic Coulomb interaction
energy. When this quantity is subtracted from the CMRT total energy, the
equilibrium lattice volume shifts to approximately that of the HF equilibrium
volume. This alignment might seem coincidental, given that CMRT and HF
converge to distinct ground states with varying orbital occupations in the
minority spin channel. Nevertheless, this shifting trend underscores the
significance of accurately addressing correlation effects for a precise
depiction of a physical system. Segmenting $E_{local}$ into two-body energy
components reveals a competition of correlation energy across different
spin-spin channels, as illustrated in Fig. \ref{eng_2D}. The dominant roles of
the electronic correlation of $t_{2g}$ and $e_{g}$ states in the minority spin
channel are further highlighted in Fig. \ref{MCFLUC}. Concurrently, these
figures emphasize the weak correlation present within the majority of spin
channels of these states---a perspective somewhat at odds with the insights
from $f\left(  z\right)  $ in Fig. \ref{Z}. One potential explanation is that
$\chi_{\alpha\sigma,\beta\sigma^{\prime}}$ provides static correlation data
for two electrons in a system's final state, which emerges after the
culmination of all inherent physical screening and damping effects. In
contrast, $f\left(  z\right)  $ may carry dynamical significance for
individual orbits, facilitating quasiparticle motion renormalization and
giving rise to necessary screening and damping effects. While $\chi
_{\alpha\sigma,\beta\sigma^{\prime}}$ could suggest the ease with which two
electrons approach each other, it doesn't necessarily correlate
straightforwardly with the single particle-related Gutzwiller renormalization
factor $f\left(  z\right)  $ under a mean field scenario. Such an
interpretation might also reconcile a statement made in a DFT+DMFT calculation
emphasizing a strong correlation effect in the majority spin
channel\cite{Grechnev2007_LDA-DMFT_QSSpectr_U2.3J0.9_PhysRevB.76.035107} by
noting an intricate connection between self-energy and Gutzwiller
renormalization
prefactor\cite{Lanata2015_LDA-G_RelatedTo_LDA-DMFT_PhysRevX.5.011008}.

Analysis of local fluctuations and pseudo-charge correlators suggested
distinct correlation patterns for 3d orbits in the majority and minority spin
channels. A closer look at pseudo-charge correlators associated with $t_{2g}$
and $e_{g}$ states indicates that both orbits exhibit significant interactions
within and among themselves in the minority spin channel, without major
qualitative differences. Hence, the approach of categorizing $t_{2g}$ and
$e_{g}$ states as purely itinerant and localized states or attributing them
different electronic
characteristics\cite{Katanin2010_LDA-DMFT_U2.3J0.9_PhysRevB.81.045117} isn't
wholly corroborated by our findings. Subsequent analysis exploring local
fluctuation was carried out. While NLCF can be insightful for analyzing
electronic localization in strongly correlated systems, it didn't yield any
substantial insights for the bulk BCC iron system. This aligns with the notion
that localization-delocalization dynamics are not a primary concern here. On
the other hand, by accessing the standardized NLCF for 3d orbits and
contrasting them with HF computations, it becomes evident that $t_{2g}$ and
$e_{g}$ states have distinct behaviors in the majority spin channel. While
they almost retain their local orbit occupations, their local charge
fluctuations are modulated in opposing directions, optimizing electronic
correlation energy for the CMRT ground state. The profound difference in the
behaviors of $t_{2g}$ and $e_{g}$ states within the majority spin channel
warrants further investigation.

\section{Summary}

In this study, we expanded the capabilities of CMRT, an entirely \textit{ab
initio} approach for correlated electron systems, to accommodate magnetization
by facilitating straightforward spin polarization within the system. We put
this formalism to the test, benchmarking it against the established
ferromagnetic system of bulk bcc Fe. Interestingly, we found that utilizing
spin-independent sum rule energy coefficients yielded the most accurate
results in the CMRT total energy computations. This finding is in harmony with
a raw \textit{ab initio} Hamiltonian employing spin-independent energy
parameters. We charted the E-V curve for this system, deriving equilibrium
attributes like volume and bulk modulus. These values align closely with
experimental data and compare positively to other ab initio methodologies.
Furthermore, our constructed P-V curve not only mirrors experimental results
but also demonstrates better concordance than GGA predictions. Diving deeper,
we extensively examined local physical metrics, encompassing local orbit
occupation, local spin and charge fluctuations, and local correlation effects
using new measures introduced in this study. Our findings pinpointed the
primary correlation impact to the 3d orbits within the minority spin channel
and highlighted subtle distinctions between $t_{2g}$ and $e_{g}$ states. While
the majority spin channel exhibits weak correlation, the behaviors of $t_{2g}$
and $e_{g}$ are notably different. This discrepancy might hinge on the method
used, and its physical implications remain unclear.

\section{Acknowledgement}

\textit{Acknowledgement} We would like to thank F. Zhang and J. H. Zhang for
valuable discussions. This work was supported by the U.S. Department of Energy
(DOE), Office of Science, Basic Energy Sciences, Materials Science and
Engineering Division, including the computer time support from the National
Energy Research Scientific Computing Center (NERSC) in Berkeley, CA. The
research was performed at Ames Laboratory, which is operated for the U.S. DOE
by Iowa State University under Contract No. DEAC02-07CH11358.
\bibliographystyle{apsrev4-2}
\bibliography{aCMRsr_BCC_Fe_short}

%apsrev4-2.bst 2019-01-14 (MD) hand-edited version of apsrev4-1.bst
%Control: key (0)
%Control: author (72) initials jnrlst
%Control: editor formatted (1) identically to author
%Control: production of article title (-1) disabled
%Control: page (0) single
%Control: year (1) truncated
%Control: production of eprint (0) enabled
\begin{thebibliography}{45}%
\makeatletter
\providecommand \@ifxundefined [1]{%
 \@ifx{#1\undefined}
}%
\providecommand \@ifnum [1]{%
 \ifnum #1\expandafter \@firstoftwo
 \else \expandafter \@secondoftwo
 \fi
}%
\providecommand \@ifx [1]{%
 \ifx #1\expandafter \@firstoftwo
 \else \expandafter \@secondoftwo
 \fi
}%
\providecommand \natexlab [1]{#1}%
\providecommand \enquote  [1]{``#1''}%
\providecommand \bibnamefont  [1]{#1}%
\providecommand \bibfnamefont [1]{#1}%
\providecommand \citenamefont [1]{#1}%
\providecommand \href@noop [0]{\@secondoftwo}%
\providecommand \href [0]{\begingroup \@sanitize@url \@href}%
\providecommand \@href[1]{\@@startlink{#1}\@@href}%
\providecommand \@@href[1]{\endgroup#1\@@endlink}%
\providecommand \@sanitize@url [0]{\catcode `\\12\catcode `\$12\catcode
  `\&12\catcode `\#12\catcode `\^12\catcode `\_12\catcode `\%12\relax}%
\providecommand \@@startlink[1]{}%
\providecommand \@@endlink[0]{}%
\providecommand \url  [0]{\begingroup\@sanitize@url \@url }%
\providecommand \@url [1]{\endgroup\@href {#1}{\urlprefix }}%
\providecommand \urlprefix  [0]{URL }%
\providecommand \Eprint [0]{\href }%
\providecommand \doibase [0]{https://doi.org/}%
\providecommand \selectlanguage [0]{\@gobble}%
\providecommand \bibinfo  [0]{\@secondoftwo}%
\providecommand \bibfield  [0]{\@secondoftwo}%
\providecommand \translation [1]{[#1]}%
\providecommand \BibitemOpen [0]{}%
\providecommand \bibitemStop [0]{}%
\providecommand \bibitemNoStop [0]{.\EOS\space}%
\providecommand \EOS [0]{\spacefactor3000\relax}%
\providecommand \BibitemShut  [1]{\csname bibitem#1\endcsname}%
\let\auto@bib@innerbib\@empty
%</preamble>
\bibitem [{\citenamefont {Kittel}(2004)}]{IntroToSolidStatePhys_Kittel2004}%
  \BibitemOpen
  \bibfield  {author} {\bibinfo {author} {\bibfnamefont {C.}~\bibnamefont
  {Kittel}},\ }\href@noop {} {\emph {\bibinfo {title} {Introduction to Solid
  State Physics}}},\ \bibinfo {edition} {8th}\ ed.\ (\bibinfo  {publisher}
  {Wiley},\ \bibinfo {year} {2004})\BibitemShut {NoStop}%
\bibitem [{\citenamefont {Kanematsu}\ \emph {et~al.}(2017)\citenamefont
  {Kanematsu}, \citenamefont {Misawa}, \citenamefont {Shiga}, \citenamefont
  {Wada},\ and\ \citenamefont {Wijn}}]{LandoltBornstein2017}%
  \BibitemOpen
  \bibfield  {author} {\bibinfo {author} {\bibfnamefont {K.}~\bibnamefont
  {Kanematsu}}, \bibinfo {author} {\bibfnamefont {S.}~\bibnamefont {Misawa}},
  \bibinfo {author} {\bibfnamefont {M.}~\bibnamefont {Shiga}}, \bibinfo
  {author} {\bibfnamefont {H.}~\bibnamefont {Wada}},\ and\ \bibinfo {author}
  {\bibfnamefont {H.}~\bibnamefont {Wijn}},\ }\href@noop {} {\emph {\bibinfo
  {title} {Landolt-Bornstein: Numerical data and functional relationships in
  science and technology}}},\ edited by\ \bibinfo {editor} {\bibfnamefont
  {W.}~\bibnamefont {Martienssen}},\ \bibinfo {series} {New series},
  Vol.~\bibinfo {volume} {32}\ (\bibinfo  {publisher} {Springer},\ \bibinfo
  {year} {2017})\ \bibinfo {note} {group III: Condensed Matter Volume 32;
  Magnetic Properties of Metals, Supplement to Volume 19, Subvolume A, 3d, 4d
  and 5d Elements, Alloys and Compounds}\BibitemShut {NoStop}%
\bibitem [{\citenamefont {Antonides}\ \emph {et~al.}(1977)\citenamefont
  {Antonides}, \citenamefont {Jose},\ and\ \citenamefont
  {Sewatzky}}]{Antonides1977_Auger_Ueff_Fe}%
  \BibitemOpen
  \bibfield  {author} {\bibinfo {author} {\bibfnamefont {E.}~\bibnamefont
  {Antonides}}, \bibinfo {author} {\bibfnamefont {E.~C.}\ \bibnamefont
  {Jose}},\ and\ \bibinfo {author} {\bibfnamefont {G.~A.}\ \bibnamefont
  {Sewatzky}},\ }\href@noop {} {\bibfield  {journal} {\bibinfo  {journal}
  {Physical Review B}\ }\textbf {\bibinfo {volume} {15}},\ \bibinfo {pages}
  {1669} (\bibinfo {year} {1977})}\BibitemShut {NoStop}%
\bibitem [{\citenamefont {Yin}\ \emph {et~al.}(1977)\citenamefont {Yin},
  \citenamefont {Tsang},\ and\ \citenamefont
  {Adler}}]{Yin1977_Auger_delocal_Ueff_PhysRevB.15.2974}%
  \BibitemOpen
  \bibfield  {author} {\bibinfo {author} {\bibfnamefont {L.~I.}\ \bibnamefont
  {Yin}}, \bibinfo {author} {\bibfnamefont {T.}~\bibnamefont {Tsang}},\ and\
  \bibinfo {author} {\bibfnamefont {I.}~\bibnamefont {Adler}},\ }\href
  {https://doi.org/10.1103/PhysRevB.15.2974} {\bibfield  {journal} {\bibinfo
  {journal} {Phys. Rev. B}\ }\textbf {\bibinfo {volume} {15}},\ \bibinfo
  {pages} {2974} (\bibinfo {year} {1977})}\BibitemShut {NoStop}%
\bibitem [{\citenamefont {Chandesris}\ \emph {et~al.}(1983)\citenamefont
  {Chandesris}, \citenamefont {Lecante},\ and\ \citenamefont
  {Petroff}}]{Chandesris1983_ARPES_Fe-Satellite}%
  \BibitemOpen
  \bibfield  {author} {\bibinfo {author} {\bibfnamefont {D.}~\bibnamefont
  {Chandesris}}, \bibinfo {author} {\bibfnamefont {J.}~\bibnamefont
  {Lecante}},\ and\ \bibinfo {author} {\bibfnamefont {Y.}~\bibnamefont
  {Petroff}},\ }\href@noop {} {\bibfield  {journal} {\bibinfo  {journal}
  {Physical Review B}\ }\textbf {\bibinfo {volume} {27}},\ \bibinfo {pages}
  {2630} (\bibinfo {year} {1983})}\BibitemShut {NoStop}%
\bibitem [{\citenamefont {\ifmmode \mbox{\c{S}}\else \c{S}\fi{}a\ifmmode
  \mbox{\c{s}}\else \c{s}\fi{}\ifmmode \imath \else \i
  \fi{}o\ifmmode~\breve{g}\else \u{g}\fi{}lu}\ \emph
  {et~al.}(2011)\citenamefont {\ifmmode \mbox{\c{S}}\else \c{S}\fi{}a\ifmmode
  \mbox{\c{s}}\else \c{s}\fi{}\ifmmode \imath \else \i
  \fi{}o\ifmmode~\breve{g}\else \u{g}\fi{}lu}, \citenamefont {Friedrich},\ and\
  \citenamefont
  {Bl\"ugel}}]{Sasioglu2011_cLDA_ScreenU_fullyScreenU_PhysRevB.83.121101}%
  \BibitemOpen
  \bibfield  {author} {\bibinfo {author} {\bibfnamefont {E.}~\bibnamefont
  {\ifmmode \mbox{\c{S}}\else \c{S}\fi{}a\ifmmode \mbox{\c{s}}\else
  \c{s}\fi{}\ifmmode \imath \else \i \fi{}o\ifmmode~\breve{g}\else
  \u{g}\fi{}lu}}, \bibinfo {author} {\bibfnamefont {C.}~\bibnamefont
  {Friedrich}},\ and\ \bibinfo {author} {\bibfnamefont {S.}~\bibnamefont
  {Bl\"ugel}},\ }\href {https://doi.org/10.1103/PhysRevB.83.121101} {\bibfield
  {journal} {\bibinfo  {journal} {Phys. Rev. B}\ }\textbf {\bibinfo {volume}
  {83}},\ \bibinfo {pages} {121101} (\bibinfo {year} {2011})}\BibitemShut
  {NoStop}%
\bibitem [{\citenamefont {Guti\'errez}\ and\ \citenamefont
  {L\'opez}(1997)}]{Gutierrez1997_ARPES_Fe-Satellite_PhysRevB.56.1111}%
  \BibitemOpen
  \bibfield  {author} {\bibinfo {author} {\bibfnamefont {A.}~\bibnamefont
  {Guti\'errez}}\ and\ \bibinfo {author} {\bibfnamefont {M.~F.}\ \bibnamefont
  {L\'opez}},\ }\href {https://doi.org/10.1103/PhysRevB.56.1111} {\bibfield
  {journal} {\bibinfo  {journal} {Phys. Rev. B}\ }\textbf {\bibinfo {volume}
  {56}},\ \bibinfo {pages} {1111} (\bibinfo {year} {1997})}\BibitemShut
  {NoStop}%
\bibitem [{\citenamefont {Wohlfarth}(1980)}]{Wohlfarth1980_Fe-Ni_co_basic}%
  \BibitemOpen
  \bibfield  {author} {\bibinfo {author} {\bibfnamefont {E.}~\bibnamefont
  {Wohlfarth}},\ }\bibinfo {title} {Iron, cobalt and nickel},\ in\ \href@noop
  {} {\emph {\bibinfo {booktitle} {Handbook of Magnetic Materials}}},\
  Vol.~\bibinfo {volume} {1},\ \bibinfo {editor} {edited by\ \bibinfo {editor}
  {\bibfnamefont {E.}~\bibnamefont {Wohlfarth}}}\ (\bibinfo  {publisher}
  {North-Holland Publishing Company},\ \bibinfo {year} {1980})\ p.~\bibinfo
  {pages} {1}\BibitemShut {NoStop}%
\bibitem [{\citenamefont {Staunton}\ \emph {et~al.}(1984)\citenamefont
  {Staunton}, \citenamefont {Gy\"orffy}, \citenamefont {Pindor}, \citenamefont
  {Stocks},\ and\ \citenamefont {Winter}}]{Staunton1984_DLM_JMMM.45.15}%
  \BibitemOpen
  \bibfield  {author} {\bibinfo {author} {\bibfnamefont {J.}~\bibnamefont
  {Staunton}}, \bibinfo {author} {\bibfnamefont {B.}~\bibnamefont {Gy\"orffy}},
  \bibinfo {author} {\bibfnamefont {A.}~\bibnamefont {Pindor}}, \bibinfo
  {author} {\bibfnamefont {G.}~\bibnamefont {Stocks}},\ and\ \bibinfo {author}
  {\bibfnamefont {H.}~\bibnamefont {Winter}},\ }\href@noop {} {\bibfield
  {journal} {\bibinfo  {journal} {Journal of Magnetism and Magnetic Materials}\
  }\textbf {\bibinfo {volume} {45}},\ \bibinfo {pages} {15} (\bibinfo {year}
  {1984})}\BibitemShut {NoStop}%
\bibitem [{\citenamefont {Stearns}(1973)}]{Stearns1973_95Pecent_LocalMom_PRB}%
  \BibitemOpen
  \bibfield  {author} {\bibinfo {author} {\bibfnamefont {M.~B.}\ \bibnamefont
  {Stearns}},\ }\href@noop {} {\bibfield  {journal} {\bibinfo  {journal}
  {Physical Review B}\ }\textbf {\bibinfo {volume} {8}} (\bibinfo {year}
  {1973})}\BibitemShut {NoStop}%
\bibitem [{\citenamefont {Singh}\ \emph {et~al.}(1991)\citenamefont {Singh},
  \citenamefont {Pickett},\ and\ \citenamefont
  {Krakauer}}]{Singh1991_GGA_Fe_DOSatEf}%
  \BibitemOpen
  \bibfield  {author} {\bibinfo {author} {\bibfnamefont {D.~J.}\ \bibnamefont
  {Singh}}, \bibinfo {author} {\bibfnamefont {W.~E.}\ \bibnamefont {Pickett}},\
  and\ \bibinfo {author} {\bibfnamefont {H.}~\bibnamefont {Krakauer}},\
  }\href@noop {} {\bibfield  {journal} {\bibinfo  {journal} {Physical Review
  B}\ }\textbf {\bibinfo {volume} {43}},\ \bibinfo {pages} {11628} (\bibinfo
  {year} {1991})}\BibitemShut {NoStop}%
\bibitem [{\citenamefont {Katanin}\ \emph {et~al.}(2010)\citenamefont
  {Katanin}, \citenamefont {Poteryaev}, \citenamefont {Efremov}, \citenamefont
  {Shorikov}, \citenamefont {Skornyakov}, \citenamefont {Korotin},\ and\
  \citenamefont {Anisimov}}]{Katanin2010_LDA-DMFT_U2.3J0.9_PhysRevB.81.045117}%
  \BibitemOpen
  \bibfield  {author} {\bibinfo {author} {\bibfnamefont {A.~A.}\ \bibnamefont
  {Katanin}}, \bibinfo {author} {\bibfnamefont {A.~I.}\ \bibnamefont
  {Poteryaev}}, \bibinfo {author} {\bibfnamefont {A.~V.}\ \bibnamefont
  {Efremov}}, \bibinfo {author} {\bibfnamefont {A.~O.}\ \bibnamefont
  {Shorikov}}, \bibinfo {author} {\bibfnamefont {S.~L.}\ \bibnamefont
  {Skornyakov}}, \bibinfo {author} {\bibfnamefont {M.~A.}\ \bibnamefont
  {Korotin}},\ and\ \bibinfo {author} {\bibfnamefont {V.~I.}\ \bibnamefont
  {Anisimov}},\ }\href@noop {} {\bibfield  {journal} {\bibinfo  {journal}
  {Phys. Rev. B}\ }\textbf {\bibinfo {volume} {81}},\ \bibinfo {pages} {045117}
  (\bibinfo {year} {2010})}\BibitemShut {NoStop}%
\bibitem [{\citenamefont {Wang}\ \emph {et~al.}(1985)\citenamefont {Wang},
  \citenamefont {Klein},\ and\ \citenamefont
  {Krakauer}}]{Wang1985_LSDA_wrongGS}%
  \BibitemOpen
  \bibfield  {author} {\bibinfo {author} {\bibfnamefont {C.~S.}\ \bibnamefont
  {Wang}}, \bibinfo {author} {\bibfnamefont {B.~M.}\ \bibnamefont {Klein}},\
  and\ \bibinfo {author} {\bibfnamefont {H.}~\bibnamefont {Krakauer}},\
  }\href@noop {} {\bibfield  {journal} {\bibinfo  {journal} {Physical Review
  Letters}\ }\textbf {\bibinfo {volume} {54}},\ \bibinfo {pages} {1852}
  (\bibinfo {year} {1985})}\BibitemShut {NoStop}%
\bibitem [{\citenamefont {Stixrude}\ \emph {et~al.}(1994)\citenamefont
  {Stixrude}, \citenamefont {Cohen},\ and\ \citenamefont
  {Singh}}]{Stixrude1994_GGA_Fe_PV}%
  \BibitemOpen
  \bibfield  {author} {\bibinfo {author} {\bibfnamefont {L.}~\bibnamefont
  {Stixrude}}, \bibinfo {author} {\bibfnamefont {R.~E.}\ \bibnamefont
  {Cohen}},\ and\ \bibinfo {author} {\bibfnamefont {D.~J.}\ \bibnamefont
  {Singh}},\ }\href@noop {} {\bibfield  {journal} {\bibinfo  {journal}
  {Physical Review B}\ }\textbf {\bibinfo {volume} {50}},\ \bibinfo {pages}
  {6442} (\bibinfo {year} {1994})}\BibitemShut {NoStop}%
\bibitem [{\citenamefont {Kresse}\ and\ \citenamefont
  {Joubert}(1999)}]{Kresse1999_PRB_Fe_LDA-PAW}%
  \BibitemOpen
  \bibfield  {author} {\bibinfo {author} {\bibfnamefont {G.}~\bibnamefont
  {Kresse}}\ and\ \bibinfo {author} {\bibfnamefont {D.}~\bibnamefont
  {Joubert}},\ }\href@noop {} {\bibfield  {journal} {\bibinfo  {journal}
  {Physical Review B}\ }\textbf {\bibinfo {volume} {59}} (\bibinfo {year}
  {1999})}\BibitemShut {NoStop}%
\bibitem [{\citenamefont {Schafer}\ \emph {et~al.}(2005)\citenamefont
  {Schafer}, \citenamefont {Hoinkis}, \citenamefont {Rotenberg}, \citenamefont
  {Blaha},\ and\ \citenamefont {Claessen}}]{schafer2005_ARPES_FSContour}%
  \BibitemOpen
  \bibfield  {author} {\bibinfo {author} {\bibfnamefont {J.}~\bibnamefont
  {Schafer}}, \bibinfo {author} {\bibfnamefont {M.}~\bibnamefont {Hoinkis}},
  \bibinfo {author} {\bibfnamefont {E.}~\bibnamefont {Rotenberg}}, \bibinfo
  {author} {\bibfnamefont {P.}~\bibnamefont {Blaha}},\ and\ \bibinfo {author}
  {\bibfnamefont {R.}~\bibnamefont {Claessen}},\ }\href@noop {} {\bibfield
  {journal} {\bibinfo  {journal} {Physical Review B}\ }\textbf {\bibinfo
  {volume} {72}} (\bibinfo {year} {2005})}\BibitemShut {NoStop}%
\bibitem [{\citenamefont
  {Gunnarsson}(1976)}]{Gunnarsson1976_StonerCrit-LDA_JPF}%
  \BibitemOpen
  \bibfield  {author} {\bibinfo {author} {\bibfnamefont {O.}~\bibnamefont
  {Gunnarsson}},\ }\href@noop {} {\bibfield  {journal} {\bibinfo  {journal} {J.
  Phys. F: Met. Phys.}\ }\textbf {\bibinfo {volume} {6}},\ \bibinfo {pages}
  {587} (\bibinfo {year} {1976})}\BibitemShut {NoStop}%
\bibitem [{\citenamefont {Yamasaki}\ and\ \citenamefont
  {Fujiwara}(2003)}]{Yamasaki2003_GW}%
  \BibitemOpen
  \bibfield  {author} {\bibinfo {author} {\bibfnamefont {A.}~\bibnamefont
  {Yamasaki}}\ and\ \bibinfo {author} {\bibfnamefont {T.}~\bibnamefont
  {Fujiwara}},\ }\href@noop {} {\bibfield  {journal} {\bibinfo  {journal} {J.
  Phys. Soc. Jpn.}\ }\textbf {\bibinfo {volume} {72}},\ \bibinfo {pages} {607}
  (\bibinfo {year} {2003})}\BibitemShut {NoStop}%
\bibitem [{\citenamefont {Kotani}\ \emph {et~al.}(2007)\citenamefont {Kotani},
  \citenamefont {Schilfgaarde},\ and\ \citenamefont
  {Faleev}}]{Kotani2007_QSGW}%
  \BibitemOpen
  \bibfield  {author} {\bibinfo {author} {\bibfnamefont {T.}~\bibnamefont
  {Kotani}}, \bibinfo {author} {\bibfnamefont {M.~v.}\ \bibnamefont
  {Schilfgaarde}},\ and\ \bibinfo {author} {\bibfnamefont {S.~V.}\ \bibnamefont
  {Faleev}},\ }\href@noop {} {\bibfield  {journal} {\bibinfo  {journal}
  {Physical Review B}\ }\textbf {\bibinfo {volume} {76}} (\bibinfo {year}
  {2007})}\BibitemShut {NoStop}%
\bibitem [{\citenamefont {Sponza}\ \emph {et~al.}(2017)\citenamefont {Sponza},
  \citenamefont {Pisanti}, \citenamefont {Vishina}, \citenamefont {Pashov},
  \citenamefont {Weber}, \citenamefont {van Schilfgaarde}, \citenamefont
  {Acharya}, \citenamefont {Vidal},\ and\ \citenamefont
  {Kotliar}}]{Sponza2017_QSGW_Fe-Ni_PhysRevB.95.041112}%
  \BibitemOpen
  \bibfield  {author} {\bibinfo {author} {\bibfnamefont {L.}~\bibnamefont
  {Sponza}}, \bibinfo {author} {\bibfnamefont {P.}~\bibnamefont {Pisanti}},
  \bibinfo {author} {\bibfnamefont {A.}~\bibnamefont {Vishina}}, \bibinfo
  {author} {\bibfnamefont {D.}~\bibnamefont {Pashov}}, \bibinfo {author}
  {\bibfnamefont {C.}~\bibnamefont {Weber}}, \bibinfo {author} {\bibfnamefont
  {M.}~\bibnamefont {van Schilfgaarde}}, \bibinfo {author} {\bibfnamefont
  {S.}~\bibnamefont {Acharya}}, \bibinfo {author} {\bibfnamefont
  {J.}~\bibnamefont {Vidal}},\ and\ \bibinfo {author} {\bibfnamefont
  {G.}~\bibnamefont {Kotliar}},\ }\href
  {https://doi.org/10.1103/PhysRevB.95.041112} {\bibfield  {journal} {\bibinfo
  {journal} {Phys. Rev. B}\ }\textbf {\bibinfo {volume} {95}},\ \bibinfo
  {pages} {041112} (\bibinfo {year} {2017})}\BibitemShut {NoStop}%
\bibitem [{\citenamefont {Barreteau}\ \emph {et~al.}(2004)\citenamefont
  {Barreteau}, \citenamefont {Desjonqu\`eres}, \citenamefont
  {Ole\ifmmode~\acute{s}\else \'{s}\fi{}},\ and\ \citenamefont
  {Spanjaard}}]{Barreteau2004_semiHF_Good-BandStr_Fe-Co-Ni_PhysRevB.69.064432}%
  \BibitemOpen
  \bibfield  {author} {\bibinfo {author} {\bibfnamefont {C.}~\bibnamefont
  {Barreteau}}, \bibinfo {author} {\bibfnamefont {M.-C.}\ \bibnamefont
  {Desjonqu\`eres}}, \bibinfo {author} {\bibfnamefont {A.~M.}\ \bibnamefont
  {Ole\ifmmode~\acute{s}\else \'{s}\fi{}}},\ and\ \bibinfo {author}
  {\bibfnamefont {D.}~\bibnamefont {Spanjaard}},\ }\href
  {https://doi.org/10.1103/PhysRevB.69.064432} {\bibfield  {journal} {\bibinfo
  {journal} {Phys. Rev. B}\ }\textbf {\bibinfo {volume} {69}},\ \bibinfo
  {pages} {064432} (\bibinfo {year} {2004})}\BibitemShut {NoStop}%
\bibitem [{\citenamefont {Belozerov}\ and\ \citenamefont
  {Anisimov}(2014{\natexlab{a}})}]{Belozerov2014_LDA-DMFT_U4J0.9_UJvalue}%
  \BibitemOpen
  \bibfield  {author} {\bibinfo {author} {\bibfnamefont {A.~S.}\ \bibnamefont
  {Belozerov}}\ and\ \bibinfo {author} {\bibfnamefont {V.~I.}\ \bibnamefont
  {Anisimov}},\ }\href {https://doi.org/10.1088/0953-8984/26/37/375601}
  {\bibfield  {journal} {\bibinfo  {journal} {J. Phys.: Condens. Matter}\
  }\textbf {\bibinfo {volume} {26}},\ \bibinfo {pages} {375601} (\bibinfo
  {year} {2014}{\natexlab{a}})}\BibitemShut {NoStop}%
\bibitem [{\citenamefont {S\'anchez-Barriga}\ \emph {et~al.}(2009)\citenamefont
  {S\'anchez-Barriga}, \citenamefont {Fink}, \citenamefont {Boni},
  \citenamefont {Di~Marco}, \citenamefont {Braun}, \citenamefont {Min\'ar},
  \citenamefont {Varykhalov}, \citenamefont {Rader}, \citenamefont {Bellini},
  \citenamefont {Manghi}, \citenamefont {Ebert}, \citenamefont {Katsnelson},
  \citenamefont {Lichtenstein}, \citenamefont {Eriksson}, \citenamefont
  {Eberhardt},\ and\ \citenamefont
  {D\"urr}}]{Sanchez2009_ARPES_LDA-DMFT_U1.5J0.9_PhysRevLett.103.267203}%
  \BibitemOpen
  \bibfield  {author} {\bibinfo {author} {\bibfnamefont {J.}~\bibnamefont
  {S\'anchez-Barriga}}, \bibinfo {author} {\bibfnamefont {J.}~\bibnamefont
  {Fink}}, \bibinfo {author} {\bibfnamefont {V.}~\bibnamefont {Boni}}, \bibinfo
  {author} {\bibfnamefont {I.}~\bibnamefont {Di~Marco}}, \bibinfo {author}
  {\bibfnamefont {J.}~\bibnamefont {Braun}}, \bibinfo {author} {\bibfnamefont
  {J.}~\bibnamefont {Min\'ar}}, \bibinfo {author} {\bibfnamefont
  {A.}~\bibnamefont {Varykhalov}}, \bibinfo {author} {\bibfnamefont
  {O.}~\bibnamefont {Rader}}, \bibinfo {author} {\bibfnamefont
  {V.}~\bibnamefont {Bellini}}, \bibinfo {author} {\bibfnamefont
  {F.}~\bibnamefont {Manghi}}, \bibinfo {author} {\bibfnamefont
  {H.}~\bibnamefont {Ebert}}, \bibinfo {author} {\bibfnamefont {M.~I.}\
  \bibnamefont {Katsnelson}}, \bibinfo {author} {\bibfnamefont {A.~I.}\
  \bibnamefont {Lichtenstein}}, \bibinfo {author} {\bibfnamefont
  {O.}~\bibnamefont {Eriksson}}, \bibinfo {author} {\bibfnamefont
  {W.}~\bibnamefont {Eberhardt}},\ and\ \bibinfo {author} {\bibfnamefont
  {H.~A.}\ \bibnamefont {D\"urr}},\ }\href
  {https://doi.org/10.1103/PhysRevLett.103.267203} {\bibfield  {journal}
  {\bibinfo  {journal} {Phys. Rev. Lett.}\ }\textbf {\bibinfo {volume} {103}},\
  \bibinfo {pages} {267203} (\bibinfo {year} {2009})}\BibitemShut {NoStop}%
\bibitem [{\citenamefont {Lichtenstein}\ \emph {et~al.}(2001)\citenamefont
  {Lichtenstein}, \citenamefont {Katsnelson},\ and\ \citenamefont
  {Kotliar}}]{Lichtenstein2001_LDA-DMFT_U2.3J0.9}%
  \BibitemOpen
  \bibfield  {author} {\bibinfo {author} {\bibfnamefont {A.~I.}\ \bibnamefont
  {Lichtenstein}}, \bibinfo {author} {\bibfnamefont {M.~I.}\ \bibnamefont
  {Katsnelson}},\ and\ \bibinfo {author} {\bibfnamefont {G.}~\bibnamefont
  {Kotliar}},\ }\href@noop {} {\bibfield  {journal} {\bibinfo  {journal}
  {Physical Review Letters}\ }\textbf {\bibinfo {volume} {87}},\ \bibinfo
  {pages} {067205} (\bibinfo {year} {2001})}\BibitemShut {NoStop}%
\bibitem [{\citenamefont {Pourovskii}\ \emph {et~al.}(2014)\citenamefont
  {Pourovskii}, \citenamefont {Mravlje}, \citenamefont {Ferrero}, \citenamefont
  {Parcollet},\ and\ \citenamefont
  {Abrikosov}}]{Pourovskii2014_LDA-DMFT_U4.3J1.0_EV-PV_PhysRevB.90.155120}%
  \BibitemOpen
  \bibfield  {author} {\bibinfo {author} {\bibfnamefont {L.~V.}\ \bibnamefont
  {Pourovskii}}, \bibinfo {author} {\bibfnamefont {J.}~\bibnamefont {Mravlje}},
  \bibinfo {author} {\bibfnamefont {M.}~\bibnamefont {Ferrero}}, \bibinfo
  {author} {\bibfnamefont {O.}~\bibnamefont {Parcollet}},\ and\ \bibinfo
  {author} {\bibfnamefont {I.~A.}\ \bibnamefont {Abrikosov}},\ }\href
  {https://doi.org/10.1103/PhysRevB.90.155120} {\bibfield  {journal} {\bibinfo
  {journal} {Phys. Rev. B}\ }\textbf {\bibinfo {volume} {90}},\ \bibinfo
  {pages} {155120} (\bibinfo {year} {2014})}\BibitemShut {NoStop}%
\bibitem [{\citenamefont {Cococcioni}\ and\ \citenamefont
  {de~Gironcoli}(2005)}]{Cococcioni2005_LDAU_U2forFe_PhysRevB.71.035105}%
  \BibitemOpen
  \bibfield  {author} {\bibinfo {author} {\bibfnamefont {M.}~\bibnamefont
  {Cococcioni}}\ and\ \bibinfo {author} {\bibfnamefont {S.}~\bibnamefont
  {de~Gironcoli}},\ }\href {https://doi.org/10.1103/PhysRevB.71.035105}
  {\bibfield  {journal} {\bibinfo  {journal} {Phys. Rev. B}\ }\textbf {\bibinfo
  {volume} {71}},\ \bibinfo {pages} {035105} (\bibinfo {year}
  {2005})}\BibitemShut {NoStop}%
\bibitem [{\citenamefont {Katsnelson}\ and\ \citenamefont
  {Lichtenstein}(1999)}]{Katsnelson1999_DMFT-FLEX_U2.3J0.9}%
  \BibitemOpen
  \bibfield  {author} {\bibinfo {author} {\bibfnamefont {M.~I.}\ \bibnamefont
  {Katsnelson}}\ and\ \bibinfo {author} {\bibfnamefont {A.~I.}\ \bibnamefont
  {Lichtenstein}},\ }\href@noop {} {\bibfield  {journal} {\bibinfo  {journal}
  {Journal of Physics: Condensed Matter}\ }\textbf {\bibinfo {volume} {11}},\
  \bibinfo {pages} {1037} (\bibinfo {year} {1999})}\BibitemShut {NoStop}%
\bibitem [{\citenamefont {Anisimov}\ \emph {et~al.}(2012)\citenamefont
  {Anisimov}, \citenamefont {Belozerov}, \citenamefont {Poteryaev},\ and\
  \citenamefont {Leonov}}]{Anisimov2012_LDA-DMFT_U2.3J0.9_PhysRevB.86.035152}%
  \BibitemOpen
  \bibfield  {author} {\bibinfo {author} {\bibfnamefont {V.~I.}\ \bibnamefont
  {Anisimov}}, \bibinfo {author} {\bibfnamefont {A.~S.}\ \bibnamefont
  {Belozerov}}, \bibinfo {author} {\bibfnamefont {A.~I.}\ \bibnamefont
  {Poteryaev}},\ and\ \bibinfo {author} {\bibfnamefont {I.}~\bibnamefont
  {Leonov}},\ }\href {https://doi.org/10.1103/PhysRevB.86.035152} {\bibfield
  {journal} {\bibinfo  {journal} {Phys. Rev. B}\ }\textbf {\bibinfo {volume}
  {86}},\ \bibinfo {pages} {035152} (\bibinfo {year} {2012})}\BibitemShut
  {NoStop}%
\bibitem [{\citenamefont {Deng}\ \emph {et~al.}(2008)\citenamefont {Deng},
  \citenamefont {Dai},\ and\ \citenamefont {Fang}}]{Deng2008_GLDA_FeU7J1}%
  \BibitemOpen
  \bibfield  {author} {\bibinfo {author} {\bibfnamefont {X.}~\bibnamefont
  {Deng}}, \bibinfo {author} {\bibfnamefont {X.}~\bibnamefont {Dai}},\ and\
  \bibinfo {author} {\bibfnamefont {Z.}~\bibnamefont {Fang}},\ }\href
  {https://doi.org/10.1209/0295-5075/83/37008} {\bibfield  {journal} {\bibinfo
  {journal} {EPL (Europhysics Letters)}\ }\textbf {\bibinfo {volume} {83}},\
  \bibinfo {pages} {37008} (\bibinfo {year} {2008})}\BibitemShut {NoStop}%
\bibitem [{\citenamefont {Borghi}\ \emph {et~al.}(2014)\citenamefont {Borghi},
  \citenamefont {Fabrizio},\ and\ \citenamefont
  {Tosatti}}]{Borghi2014_PRB_GLDA_U2.5J1.2}%
  \BibitemOpen
  \bibfield  {author} {\bibinfo {author} {\bibfnamefont {G.}~\bibnamefont
  {Borghi}}, \bibinfo {author} {\bibfnamefont {M.}~\bibnamefont {Fabrizio}},\
  and\ \bibinfo {author} {\bibfnamefont {E.}~\bibnamefont {Tosatti}},\
  }\href@noop {} {\bibfield  {journal} {\bibinfo  {journal} {Physical Review
  B}\ }\textbf {\bibinfo {volume} {90}} (\bibinfo {year} {2014})}\BibitemShut
  {NoStop}%
\bibitem [{\citenamefont {Schickling}\ \emph {et~al.}(2016)\citenamefont
  {Schickling}, \citenamefont {B\"unemann}, \citenamefont {Gebhard},\ and\
  \citenamefont {Boeri}}]{Schickling2016_GLDA-U9J0.54}%
  \BibitemOpen
  \bibfield  {author} {\bibinfo {author} {\bibfnamefont {T.}~\bibnamefont
  {Schickling}}, \bibinfo {author} {\bibfnamefont {J.}~\bibnamefont
  {B\"unemann}}, \bibinfo {author} {\bibfnamefont {F.}~\bibnamefont
  {Gebhard}},\ and\ \bibinfo {author} {\bibfnamefont {L.}~\bibnamefont
  {Boeri}},\ }\href@noop {} {\bibfield  {journal} {\bibinfo  {journal}
  {Physical Review B}\ }\textbf {\bibinfo {volume} {93}} (\bibinfo {year}
  {2016})}\BibitemShut {NoStop}%
\bibitem [{\citenamefont {Grechnev}\ \emph {et~al.}(2007)\citenamefont
  {Grechnev}, \citenamefont {Di~Marco}, \citenamefont {Katsnelson},
  \citenamefont {Lichtenstein}, \citenamefont {Wills},\ and\ \citenamefont
  {Eriksson}}]{Grechnev2007_LDA-DMFT_QSSpectr_U2.3J0.9_PhysRevB.76.035107}%
  \BibitemOpen
  \bibfield  {author} {\bibinfo {author} {\bibfnamefont {A.}~\bibnamefont
  {Grechnev}}, \bibinfo {author} {\bibfnamefont {I.}~\bibnamefont {Di~Marco}},
  \bibinfo {author} {\bibfnamefont {M.~I.}\ \bibnamefont {Katsnelson}},
  \bibinfo {author} {\bibfnamefont {A.~I.}\ \bibnamefont {Lichtenstein}},
  \bibinfo {author} {\bibfnamefont {J.}~\bibnamefont {Wills}},\ and\ \bibinfo
  {author} {\bibfnamefont {O.}~\bibnamefont {Eriksson}},\ }\href
  {https://doi.org/10.1103/PhysRevB.76.035107} {\bibfield  {journal} {\bibinfo
  {journal} {Phys. Rev. B}\ }\textbf {\bibinfo {volume} {76}},\ \bibinfo
  {pages} {035107} (\bibinfo {year} {2007})}\BibitemShut {NoStop}%
\bibitem [{\citenamefont {Kvashnin}\ \emph {et~al.}(2015)\citenamefont
  {Kvashnin}, \citenamefont {Gr\aa{}n\"as}, \citenamefont {Di~Marco},
  \citenamefont {Katsnelson}, \citenamefont {Lichtenstein},\ and\ \citenamefont
  {Eriksson}}]{Kvashnin2015_J2_LDA-DMFT_U2.3J0.9_PhysRevB.91.125133}%
  \BibitemOpen
  \bibfield  {author} {\bibinfo {author} {\bibfnamefont {Y.~O.}\ \bibnamefont
  {Kvashnin}}, \bibinfo {author} {\bibfnamefont {O.}~\bibnamefont
  {Gr\aa{}n\"as}}, \bibinfo {author} {\bibfnamefont {I.}~\bibnamefont
  {Di~Marco}}, \bibinfo {author} {\bibfnamefont {M.~I.}\ \bibnamefont
  {Katsnelson}}, \bibinfo {author} {\bibfnamefont {A.~I.}\ \bibnamefont
  {Lichtenstein}},\ and\ \bibinfo {author} {\bibfnamefont {O.}~\bibnamefont
  {Eriksson}},\ }\href {https://doi.org/10.1103/PhysRevB.91.125133} {\bibfield
  {journal} {\bibinfo  {journal} {Phys. Rev. B}\ }\textbf {\bibinfo {volume}
  {91}},\ \bibinfo {pages} {125133} (\bibinfo {year} {2015})}\BibitemShut
  {NoStop}%
\bibitem [{\citenamefont {Lanat\`a}\ \emph {et~al.}(2015)\citenamefont
  {Lanat\`a}, \citenamefont {Yao}, \citenamefont {Wang}, \citenamefont {Ho},\
  and\ \citenamefont
  {Kotliar}}]{Lanata2015_LDA-G_RelatedTo_LDA-DMFT_PhysRevX.5.011008}%
  \BibitemOpen
  \bibfield  {author} {\bibinfo {author} {\bibfnamefont {N.}~\bibnamefont
  {Lanat\`a}}, \bibinfo {author} {\bibfnamefont {Y.}~\bibnamefont {Yao}},
  \bibinfo {author} {\bibfnamefont {C.-Z.}\ \bibnamefont {Wang}}, \bibinfo
  {author} {\bibfnamefont {K.-M.}\ \bibnamefont {Ho}},\ and\ \bibinfo {author}
  {\bibfnamefont {G.}~\bibnamefont {Kotliar}},\ }\href
  {https://doi.org/10.1103/PhysRevX.5.011008} {\bibfield  {journal} {\bibinfo
  {journal} {Phys. Rev. X}\ }\textbf {\bibinfo {volume} {5}},\ \bibinfo {pages}
  {011008} (\bibinfo {year} {2015})}\BibitemShut {NoStop}%
\bibitem [{\citenamefont {Liu}\ \emph {et~al.}(2016)\citenamefont {Liu},
  \citenamefont {Liu}, \citenamefont {Yao}, \citenamefont {Wu}, \citenamefont
  {Wang},\ and\ \citenamefont {Ho}}]{CMRsr2016_JCTC.12.4806}%
  \BibitemOpen
  \bibfield  {author} {\bibinfo {author} {\bibfnamefont {C.}~\bibnamefont
  {Liu}}, \bibinfo {author} {\bibfnamefont {J.}~\bibnamefont {Liu}}, \bibinfo
  {author} {\bibfnamefont {Y.~X.}\ \bibnamefont {Yao}}, \bibinfo {author}
  {\bibfnamefont {P.}~\bibnamefont {Wu}}, \bibinfo {author} {\bibfnamefont
  {C.~Z.}\ \bibnamefont {Wang}},\ and\ \bibinfo {author} {\bibfnamefont
  {K.~M.}\ \bibnamefont {Ho}},\ }\href@noop {} {\bibfield  {journal} {\bibinfo
  {journal} {J. Chem. Theory Comput.}\ }\textbf {\bibinfo {volume} {12}},\
  \bibinfo {pages} {4806} (\bibinfo {year} {2016})}\BibitemShut {NoStop}%
\bibitem [{\citenamefont {Zhao}\ \emph {et~al.}(2018)\citenamefont {Zhao},
  \citenamefont {Liu}, \citenamefont {Yao}, \citenamefont {Wang},\ and\
  \citenamefont {Ho}}]{CMRsr2018_PhysRevB.97.075142}%
  \BibitemOpen
  \bibfield  {author} {\bibinfo {author} {\bibfnamefont {X.}~\bibnamefont
  {Zhao}}, \bibinfo {author} {\bibfnamefont {J.}~\bibnamefont {Liu}}, \bibinfo
  {author} {\bibfnamefont {Y.-X.}\ \bibnamefont {Yao}}, \bibinfo {author}
  {\bibfnamefont {C.-Z.}\ \bibnamefont {Wang}},\ and\ \bibinfo {author}
  {\bibfnamefont {K.-M.}\ \bibnamefont {Ho}},\ }\href@noop {} {\bibfield
  {journal} {\bibinfo  {journal} {Phys. Rev. B}\ }\textbf {\bibinfo {volume}
  {97}},\ \bibinfo {pages} {075142} (\bibinfo {year} {2018})}\BibitemShut
  {NoStop}%
\bibitem [{\citenamefont {Liu}\ \emph {et~al.}(2021{\natexlab{a}})\citenamefont
  {Liu}, \citenamefont {Zhao}, \citenamefont {Yao}, \citenamefont {Wang},\ and\
  \citenamefont {Ho}}]{CMRsr2020_JPCM_33.095902_sp_formalism}%
  \BibitemOpen
  \bibfield  {author} {\bibinfo {author} {\bibfnamefont {J.}~\bibnamefont
  {Liu}}, \bibinfo {author} {\bibfnamefont {X.}~\bibnamefont {Zhao}}, \bibinfo
  {author} {\bibfnamefont {Y.}~\bibnamefont {Yao}}, \bibinfo {author}
  {\bibfnamefont {C.-Z.}\ \bibnamefont {Wang}},\ and\ \bibinfo {author}
  {\bibfnamefont {K.-M.}\ \bibnamefont {Ho}},\ }\href@noop {} {\bibfield
  {journal} {\bibinfo  {journal} {Journal of Physics: Condensed Matter}\
  }\textbf {\bibinfo {volume} {33}},\ \bibinfo {pages} {095902} (\bibinfo
  {year} {2021}{\natexlab{a}})}\BibitemShut {NoStop}%
\bibitem [{\citenamefont {Belozerov}\ and\ \citenamefont
  {Anisimov}(2014{\natexlab{b}})}]{UJ-Depnd_Belozerov_JPCM2014}%
  \BibitemOpen
  \bibfield  {author} {\bibinfo {author} {\bibfnamefont {A.~S.}\ \bibnamefont
  {Belozerov}}\ and\ \bibinfo {author} {\bibfnamefont {V.~I.}\ \bibnamefont
  {Anisimov}},\ }\href@noop {} {\bibfield  {journal} {\bibinfo  {journal} {J.
  Phys. Condens. Matter}\ }\textbf {\bibinfo {volume} {26}},\ \bibinfo {pages}
  {375601} (\bibinfo {year} {2014}{\natexlab{b}})}\BibitemShut {NoStop}%
\bibitem [{\citenamefont {Liu}\ \emph {et~al.}(2021{\natexlab{b}})\citenamefont
  {Liu}, \citenamefont {Yao}, \citenamefont {Zhang}, \citenamefont {Ho},\ and\
  \citenamefont {Wang}}]{Jun2021_CMRT-Ce_PhysRevB.104.L081113}%
  \BibitemOpen
  \bibfield  {author} {\bibinfo {author} {\bibfnamefont {J.}~\bibnamefont
  {Liu}}, \bibinfo {author} {\bibfnamefont {Y.}~\bibnamefont {Yao}}, \bibinfo
  {author} {\bibfnamefont {J.}~\bibnamefont {Zhang}}, \bibinfo {author}
  {\bibfnamefont {K.-M.}\ \bibnamefont {Ho}},\ and\ \bibinfo {author}
  {\bibfnamefont {C.-Z.}\ \bibnamefont {Wang}},\ }\href
  {https://doi.org/10.1103/PhysRevB.104.L081113} {\bibfield  {journal}
  {\bibinfo  {journal} {Phys. Rev. B}\ }\textbf {\bibinfo {volume} {104}},\
  \bibinfo {pages} {L081113} (\bibinfo {year}
  {2021}{\natexlab{b}})}\BibitemShut {NoStop}%
\bibitem [{\citenamefont {Kresse}\ and\ \citenamefont
  {Furthm\"uller}(1996)}]{VASP_PhysRevB.54.11169}%
  \BibitemOpen
  \bibfield  {author} {\bibinfo {author} {\bibfnamefont {G.}~\bibnamefont
  {Kresse}}\ and\ \bibinfo {author} {\bibfnamefont {J.}~\bibnamefont
  {Furthm\"uller}},\ }\href@noop {} {\bibfield  {journal} {\bibinfo  {journal}
  {Phys. Rev. B}\ }\textbf {\bibinfo {volume} {54}},\ \bibinfo {pages} {11169}
  (\bibinfo {year} {1996})}\BibitemShut {NoStop}%
\bibitem [{\citenamefont {Qian}\ \emph {et~al.}(2008)\citenamefont {Qian},
  \citenamefont {Li}, \citenamefont {Qi}, \citenamefont {Wang}, \citenamefont
  {Chan}, \citenamefont {Yao}, \citenamefont {Ho},\ and\ \citenamefont
  {Yip}}]{QUAMBO_PhysRevB.78.245112}%
  \BibitemOpen
  \bibfield  {author} {\bibinfo {author} {\bibfnamefont {X.}~\bibnamefont
  {Qian}}, \bibinfo {author} {\bibfnamefont {J.}~\bibnamefont {Li}}, \bibinfo
  {author} {\bibfnamefont {L.}~\bibnamefont {Qi}}, \bibinfo {author}
  {\bibfnamefont {C.-Z.}\ \bibnamefont {Wang}}, \bibinfo {author}
  {\bibfnamefont {T.-L.}\ \bibnamefont {Chan}}, \bibinfo {author}
  {\bibfnamefont {Y.-X.}\ \bibnamefont {Yao}}, \bibinfo {author} {\bibfnamefont
  {K.-M.}\ \bibnamefont {Ho}},\ and\ \bibinfo {author} {\bibfnamefont
  {S.}~\bibnamefont {Yip}},\ }\href@noop {} {\bibfield  {journal} {\bibinfo
  {journal} {Phys. Rev. B}\ }\textbf {\bibinfo {volume} {78}},\ \bibinfo
  {pages} {245112} (\bibinfo {year} {2008})}\BibitemShut {NoStop}%
\bibitem [{\citenamefont {Dewaele}\ \emph {et~al.}(2015)\citenamefont
  {Dewaele}, \citenamefont {Denoual}, \citenamefont {Anzellini}, \citenamefont
  {Occelli}, \citenamefont {Mezouar}, \citenamefont {Cordier}, \citenamefont
  {Merkel}, \citenamefont {V\'eron},\ and\ \citenamefont
  {Rausch}}]{Dewaele2015_a-e_marten}%
  \BibitemOpen
  \bibfield  {author} {\bibinfo {author} {\bibfnamefont {A.}~\bibnamefont
  {Dewaele}}, \bibinfo {author} {\bibfnamefont {C.}~\bibnamefont {Denoual}},
  \bibinfo {author} {\bibfnamefont {S.}~\bibnamefont {Anzellini}}, \bibinfo
  {author} {\bibfnamefont {F.}~\bibnamefont {Occelli}}, \bibinfo {author}
  {\bibfnamefont {M.}~\bibnamefont {Mezouar}}, \bibinfo {author} {\bibfnamefont
  {P.}~\bibnamefont {Cordier}}, \bibinfo {author} {\bibfnamefont
  {S.}~\bibnamefont {Merkel}}, \bibinfo {author} {\bibfnamefont
  {M.}~\bibnamefont {V\'eron}},\ and\ \bibinfo {author} {\bibfnamefont
  {E.}~\bibnamefont {Rausch}},\ }\href@noop {} {\bibfield  {journal} {\bibinfo
  {journal} {Physical Review B}\ }\textbf {\bibinfo {volume} {91}} (\bibinfo
  {year} {2015})}\BibitemShut {NoStop}%
\bibitem [{\citenamefont {Murnaghan}(1944)}]{Murnaghan_PNAS.30.244}%
  \BibitemOpen
  \bibfield  {author} {\bibinfo {author} {\bibfnamefont {F.~D.}\ \bibnamefont
  {Murnaghan}},\ }\href@noop {} {\bibfield  {journal} {\bibinfo  {journal}
  {Proc. Natl. Acad. Sci. U.S.A.}\ }\textbf {\bibinfo {volume} {30}},\ \bibinfo
  {pages} {244} (\bibinfo {year} {1944})}\BibitemShut {NoStop}%
\bibitem [{\citenamefont {Blundell}(2001)}]{MagnetismCondMatt_blundell2001}%
  \BibitemOpen
  \bibfield  {author} {\bibinfo {author} {\bibfnamefont {S.}~\bibnamefont
  {Blundell}},\ }\href@noop {} {\emph {\bibinfo {title} {Magnetism in Condensed
  Matter}}}\ (\bibinfo  {publisher} {Oxford University Press},\ \bibinfo
  {address} {Oxford, UK},\ \bibinfo {year} {2001})\BibitemShut {NoStop}%
\bibitem [{HF-()}]{HF-PotentialEnergy}%
  \BibitemOpen
  \href@noop {} {}\bibinfo {note} {By assuming \( n \) to be the initial charge
  in \( t_{2g} \) and supposing a subsequent change of \( x \) to produce \(
  3x/2 \) charge in the \( e_g \) states, Eq. \(\ref{localpot}\) can be
  expressed in terms of \( x \). Setting the derivative of Eq.
  \(\ref{localpot}\) to zero yields the physically relevant solution, which is,
  however, an energy maximum. Note also \( x \) is bounded between \( 0 \) and
  \( n \), which correspond to the two extreme cases.}\BibitemShut {Stop}%
\end{thebibliography}%

\end{document}